\documentclass[floatfix,twocolumn,showpacs,nofootinbib,prd,aps,tightenlines]{revtex4-1}
\usepackage{graphicx}
\usepackage{psfrag}
\usepackage{dcolumn}
\usepackage{bm}
\usepackage{amsmath}
\usepackage{bm}
\usepackage{appendix}
\usepackage[caption=false]{subfig}

\newcommand{\bea}{\begin{eqnarray}}
\newcommand{\eea}{\end{eqnarray}}
\newcommand{\beq}{\begin{equation}}
\newcommand{\eeq}{\end{equation}}

\newcommand\norm[1]{\lVert#1\rVert}

\usepackage{color}

\begin{document}

\title{Close Encounter of Three Black Holes Revisited}

\author{Alessandro Ciarfella}
\affiliation{Center for Computational Relativity and Gravitation,
School of Mathematical Sciences,
Rochester Institute of Technology, 85 Lomb Memorial Drive, Rochester,
New York 14623, USA}
\author{Giuseppe Ficarra}
\affiliation{Center for Computational Relativity and Gravitation,
School of Mathematical Sciences,
Rochester Institute of Technology, 85 Lomb Memorial Drive, Rochester,
New York 14623, USA}
\author{Carlos O. Lousto}
\affiliation{Center for Computational Relativity and Gravitation,
School of Mathematical Sciences,
Rochester Institute of Technology, 85 Lomb Memorial Drive, Rochester,
New York 14623, USA}

\date{\today}

\begin{abstract}
We study the evolution of close triple black hole system with full
numerical relativity techniques. We consider an equal mass non
spinning hierarchical system with an inner binary ten orbits away from
merger and study the effects of the third outer black hole on the
binary's merger time and its eccentricity evolution. We find a generic
time delay and an increase in the number of orbits to merger of the
binary, that can be modeled versus the distance $D$ to the
third black hole as $\sim1/D^{2.5}$.  On the other hand, we find that
the orientation of the third black hole orbit has little effect on the
binary's merger time when considering a fiducial initial distance of
$D=30M$ to the binary (with initial orbital separation $d=8M$).  In those
scenarios the evolution of the inner binary eccentricity presents a
steady decay, roughly as expected, but in addition shows a modulation
with the time scale of the outer third black hole orbital semiperiod
around the binary, resembling a beating frequency.
\end{abstract}

\maketitle

\section{Introduction}\label{sec:Intro}

Triple black hole systems have a renewed interest since the
observation that some of the gravitational waves signals detected by
the LIGO-Virgo collaboration may be the product of highly eccentric
black hole mergers \cite{Gayathri:2020coq} and that one of the
scenarios for creating those eccentricities may be the product of
three body Lidov-Kozai interactions \cite{Yu:2020iqj}.  In this
scenario, a tertiary companion on a sufficiently inclined outer orbit
could drive the inner binary to extreme eccentricities, leading to
efficient gravitational radiation and orbital decay.  Also see
\cite{DallAmico:2021umv} for a formation scenario of GW190521 via
three-body encounters in young massive star clusters.

The triple channel predicts a distinct region of the total mass, mass
ratio, and spin parameter space for merging binary black holes, which
can be used to disentangle the triple contribution to the overall
observed gravitational wave sources.  For a detailed study of the mass
ratio distribution of binary black hole mergers induced by tertiary
companions in triple systems see Ref.~\cite{Martinez:2021tmr}.  Close
encounters of stars with stellar-mass black hole binaries have been
studied in \cite{Ryu:2022qpo}.


Close encounters of three black holes require Numerical Relativity
techniques.  Full numerical evolution of triple systems are
challenging due to the need to track three black holes and the
different scales of time-integration involved in the solution.  In
Ref.~\cite{Campanelli:2007ea,Lousto:2007rj} we have performed
prototypical evolutions of such systems and evaluated its accuracy
compared to Newtonian and post-Newtonian evolutions
\cite{Lousto:2007ji}.

In this paper we will revisit this scenario and evolve triple systems
using full numerical techniques to assess the prompt or delayed merger
and eccentricity evolution of a binary in a hierarchical triple
system.

\section{Approximate initial data}\label{sec:ID}

In Ref.~\cite{Lousto:2007rj} we have performed the three black holes
prototypical studies from approximate initial data, based on
\cite{Laguna:2003sr} and extended to include terms of the sort
$\vec{S}_i\times\vec{P}_i$ representing interactions of spin with
linear momentum in an expansion to leading order on those intrinsic
parameters of the holes.  In \cite{Galaviz:2010mx} a similar study was
made using exact initial data and found (when using the same raw 3BH
parameters) some deviations in the long term evolutions when compared
to the corresponding approximate initial data.  Here we will
introduce two sets improvements to the approximate initial data for
multi black hole configurations. As already pointed out in
\cite{Lousto:2007rj}, a normalization for the parameters makes notable
improvements in, for instance, the resulting waveforms of 2BH (See
Fig. 1 in \cite{Lousto:2007rj}). To that end we will normalize data to
the initial (sum of) horizon masses as computed fully numerically.
The second improvement is to compute the next order expansion in the
solutions to the Bowen-York \cite{Bowen:1980yu} initial data set. We
will test those improvements by direct comparison with the ``exact''
initial data for 2BH.

Here we provide some details on how we find a perturbative solution of
the Hamiltonian constraint equation, since in the Bowen-York approach
\cite{Bowen:1980yu} the momentum constraint is solved exactly.  Hence,
the scope of this section is to solve perturbatively the partial
differential equation for the conformal factor $\phi$
\begin{equation}
\Delta \phi = -\frac{1}{8}\phi^{-7}\hat{A}^{ij}\hat{A}_{ij},
\end{equation}
with
\begin{eqnarray}
  \hat{A}^{ij} = \sum_a^{N_{BHs}}\left(\frac{3}{2r_a^2}\left[2P_a^{(i}n_a^{j)} + (n_a^in_a^j-\eta^{ij})P_{ak}n_a^k\right]\right.\nonumber\\
  \left.+ \frac{6}{r_a^3} n_a^{(i}\epsilon^{j)kl}J_{ak}n_{al}\right),
\end{eqnarray}
where we label the momentum and the spin of the holes as 
$\textbf{P}_i$ and $\textbf{J}_i$, following the notation of \cite{Lousto:2007rj}.

For this purpose we start from the analytical solution at order 0th given by
\begin{equation}
\phi_0 = 1 + \sum_a^{N_{BHs}}\frac{m_a}{2 r_a},
\end{equation}
which solves
\begin{equation}
\Delta \phi_0 = 0.
\end{equation}

To find the first perturbative order $u$ of the solution we consider the equation
\begin{equation}\label{Perturbation_1}
\Delta u_1 = -\frac{1}{8}\phi_0^{-7} \hat{A}^{ij}\hat{A}_{ij}.
\end{equation}

\subsection{One Black Hole}

Let us consider Eq.~(\ref{Perturbation_1}) for a single black hole. The term $\hat{A}^{ij}\hat{A}_{ij}$ is given by
\begin{equation}
\hat{A}^{ij}\hat{A}_{ij} = \frac{18 J^2}{r^6} \left(1-x_\textbf{J}^2\right)+\frac{9 P^2}{2 r^4}\left(2x_\textbf{P}^2-1\right)+\frac{18}{r^5}x_{\textbf{P} \times \textbf{J}} \norm{\textbf{P}\times \textbf{J}},
\end{equation}
where $x_\textbf{J}$, $x_\textbf{P}$ and $x_{\textbf{P}\times\textbf{J}}$ are respectively $\cos{\theta_\textbf{J}}$, $\cos{\theta_\textbf{P}}$ and $\cos{\theta_{\textbf{P}\times\textbf{J}}}$.

Since Eq.~(\ref{Perturbation_1}) is linear, the solution can be written as
\begin{equation}
u_1 = F(r,x_\textbf{P}) \textbf{P}^2 + G(r,x_\textbf{J}) \textbf{J}^2 + H(r,x_{\textbf{P}\times\textbf{J}}) \norm{\textbf{J} \times \textbf{P}}.
\end{equation}

In this way we can solve the equation separately for the functions $F(r,x_\textbf{P})$, $G(r,x_\textbf{J})$, $H(r,x_{\textbf{P}\times\textbf{J}})$.

In particular, it is convenient to write the source term in terms of Legendre polynomials. By doing so, we can solve the angular part of the equations algebraically
\begin{align}
 \begin{split}
&\Delta F(r,x_\textbf{P}) = \frac{3}{2 r^4}\left(5 P_0(x_\textbf{P}) + 4P_2(x_\textbf{P})\right),\\
&\Delta G(r,x_\textbf{J}) = \frac{12 J^2}{r^6} \left(P_0(x_\textbf{J}) - P_2(x_\textbf{J})\right),\\
&\Delta H(r,x_{\textbf{P}\times\textbf{J}}) = \frac{18}{r^5}P_1(x_{\textbf{P} \times \textbf{J}}).
 \end{split}
\end{align}

Thus the solution to first order is 
\begin{equation}
\phi_1 = \phi_0 + F(r,x_\textbf{P}) \textbf{P}^2 + G(r,x_\textbf{J}) \textbf{J}^2 + H(r,x_{\textbf{P}\times\textbf{J}}) \norm{\textbf{J} \times \textbf{P}},
\end{equation}
where the functions $F(r,x_\textbf{P})$, $ G(r,x_\textbf{J})$, $H(r,x_{\textbf{P}\times\textbf{J}})$ are explicitly given by
\begin{align}
\begin{split}
&F(r,x_\textbf{P})=\frac{1}{160 (u+1)^5} 
  [u^4 (3 x_{\textbf{P}}^2-1) (84 u^5+378 u^4\\
    &\quad\quad+658 u^3+539 u^2+192 u+84 (u+1)^5 u \log (u)\\
&\quad\quad-84 (u+1)^5 u \log (u+1)+15)]+\frac{u^2}{32} \left(1-\frac{1}{(u+1)^5}\right),\\
&G(r,x_\textbf{J})=\frac{ u^5 \left(1-3 x_{\textbf{J}}^2\right)}{40 (u+1)^5}+\frac{ \left(u^4+5 u^3+10 u^2+5 u+1\right) u^3}{40 (u+1)^5},\\
&H(r,x_{\textbf{P}\times\textbf{J}})=-\frac{ u^4 \left(u^2+5 u+10\right) x_{\textbf{P}\times\textbf{J}}}{80 (u+1)^5},
\end{split}
\end{align}
where $u=\frac{m}{2r}$ and this solution agrees with the one given in \cite{Lousto:2007rj}.

Thus when we want to solve the second order perturbation equation for a single BH we have
\begin{equation}
    \Delta \phi_2 = -\frac{1}{8} \phi_0^{-7}\left(1-7u_1\right) \hat{A}^{ij}\hat{A}_{ij},
\end{equation}
where we used the fact that $\phi_1^{-7}\approx \phi_0^{-7}(1-7u_1)$ and hence
\begin{equation}
    \Delta u_2 = \frac{7}{8} \phi_0^{-7}u_1 \hat{A}^{ij}\hat{A}_{ij}.
\end{equation}

Using the same expansion reasoning we used for the 1st order case we can write
\begin{align}\label{Perturbation_2}
\begin{split}
  u_2 =& F_{P^4}(r,x_\textbf{P})P^4 + F_{J^4}(r,x_\textbf{J})J^4 \\
  &  +F_{({\textbf{P}\times\textbf{J}})^2}(r,x_{\textbf{P}\times\textbf{J}})\norm{{\textbf{P}\times\textbf{J}}}^2 \\
  &+ F_{P({\textbf{P}\times\textbf{J}})}(r,x_\textbf{P},\phi_\textbf{P})P\norm{{\textbf{P}\times\textbf{J}}}\\
  &+F_{J({\textbf{P}\times\textbf{J}})}(r,x_\textbf{J},\phi_\textbf{J})J\norm{{\textbf{P}\times\textbf{J}}}\\
  &+F_{J^2P^2}(r,x_\textbf{P},\phi_\textbf{P})J^2P^2.\\
\end{split}
\end{align}

Note that in Eq.(\ref{Perturbation_2}) there are terms that involve
combinations of the vectors $\textbf{P}$, $\textbf{J}$,
$\textbf{P}\times\textbf{J}$. For example, when solving for the term
$F_{J({\textbf{P}\times\textbf{J}})}(r,x_\textbf{J},\phi_\textbf{J})$
we need to write $x_{\textbf{P}\times\textbf{J}}$ in terms of
$x_\textbf{J}$ and $\phi_\textbf{J}$, as described in Appendix
\ref{Rotation_Matrix}. At this point, we can solve for the single
functions in Eq.(\ref{Perturbation_2}) as we did for
Eq.(\ref{Perturbation_1}) with the only difference that 
we need a decomposition in spherical harmonics and then solve the
resulting ordinary differential equation in the variable $r$.

\subsection{Two Black Holes}

When we want to consider two BHs we proceed in the following
way. First, let us assume that our system is bound so that we can then
assume the Virial theorem to hold approximately true, $P^2 \approx
\frac{1}{r_{12}}$ where $r_{12}$ is the distance between the two
BHs. Then we treat the solution as a superposition of the
solutions for single black holes, where we also add a term due to the interaction between the two.
Under this assumptions let us solve Eq.(\ref{Perturbation_1}) for two BHs considering the perturbation of the second black hole on the first one.
In this case we have 
\begin{equation}
\frac{1}{r_2} = \frac{1}{r_{12}} + \mathcal{O}\left(\frac{1}{r_{12}}\right)^2,
\end{equation}
Hence we get
\begin{align}
\begin{split}
  \hat{A}_{ij}\hat{A}^{ij} &= \frac{18 J_1^2}{r_1^6} \left(1-x_{\textbf{J}_1}^2\right)+\frac{9 P_1^2}{2 r_1^4}\left(2x_{\textbf{P}_1}^2-1\right)\\
 & +\frac{18}{r_1^5}x_{\textbf{P}_1 \times \textbf{J}_1} \textbf{P}_1\times \textbf{J}_1
+ \mathcal{O}\left(\frac{1}{r_{12}}\right)^2
\end{split}
\end{align}
and 
\begin{equation}
\phi_0 = 1 + \frac{m_1}{2 r_1} + \frac{m_2}{2r_2} = 1 + \frac{m_1}{2 r_1} + \frac{m_2}{2r_{12}} + \mathcal{O}\left(\frac{1}{r_{12}}\right)^2,
\end{equation}
Thus we can write 
\begin{align}
\begin{split}
  \phi_0^{-7}\hat{A}_{ij}\hat{A}^{ij} &=\left(1+\frac{m_1}{2r_1}\right)^{-7} \left(\frac{18 J_1^2}{r_1^6} \left(1-x_{\textbf{J}_1}^2\right)\right.\\
&\left.  +\frac{9 P_1^2}{2 r_1^4}\left(2x_{\textbf{P}_1}^2-1\right)+\frac{18}{r_1^5}x_{\textbf{P}_1 \times \textbf{J}_1} \textbf{P}_1\times \textbf{J}_1\right)
\\
&+ \mathcal{O}\left(\frac{P^2}{r_{12}}\right)+ \mathcal{O}\left(\frac{J^2}{r_{12}}\right)+\mathcal{O}\left(\frac{\textbf{P}\times\textbf{J}}{r_{12}}\right).
\end{split}
\end{align}

As a result we can see that to first order the solution of
Eq.~(\ref{Perturbation_1}) for two BHs is the superposition of the
solutions for single black holes.

To second order we have 
\begin{align}
\begin{split}
\phi_1^{-7}\hat{A}_{ij}\hat{A}^{ij} =&\left(1+\frac{m_1}{2r_1}+\frac{m_2}{2r_2}+u_1^1 +u_1^2\right)^{-7}\\& \left[ \hat{A_1}_{ij}\hat{A_1}^{ij}+ \hat{A_2}_{ij}\hat{A_2}^{ij} + 2\hat{A_1}_{ij}\hat{A_2}^{ij}\right]\\
=&\left(1+\frac{m_1}{2r_1}\right)^{-7} \hat{A_1}_{ij}\hat{A_1}^{ij} - 7 \phi_{01}^{-7} u_1^1\hat{A_1}_{ij}\hat{A_1}^{ij}\\& -7\phi_{01}^{-7}\frac{m_2}{2r_{12}}\hat{A_1}_{ij}\hat{A_1}^{ij}
\\&+\mathcal{O}\left(\frac{P^4}{r_{12}}\right)
+\mathcal{O}\left(\frac{J^4}{r_{12}}\right)+\mathcal{O}\left(\frac{J^2P^2}{r_{12}}\right)\\
&+\mathcal{O}\left(J^6\right)+\mathcal{O}\left(P^6\right)+\mathcal{O}\left(P^5J\right)+\mathcal{O}\left(J^5P\right)\\&
+\mathcal{O}\left(J^4P^2\right)+\mathcal{O}\left(J^3P^3\right)+\mathcal{O}\left(J^2P^4\right).
\end{split}
\end{align}

In this case we considered the second black hole as a perturbation of the first one, but the opposite is also true, so that the complete source term at second order is given by

\begin{align}\label{Source term 2 BHs}
\begin{split}
\phi_1^{-7}\hat{A}_{ij}\hat{A}^{ij} =&\left(1+\frac{m_1}{2r_1}+\frac{m_2}{2r_2}+u_1^1 +u_1^2\right)^{-7}\\& \left[ \hat{A_1}_{ij}\hat{A_1}^{ij}+ \hat{A_2}_{ij}\hat{A_2}^{ij} + 2\hat{A_1}_{ij}\hat{A_2}^{ij}\right]\\
=&\left(1+\frac{m_1}{2r_1}\right)^{-7} \hat{A_1}_{ij}\hat{A_1}^{ij} - 7 \phi_{01}^{-7} u_1^1\hat{A_1}_{ij}\hat{A_1}^{ij}\\& -7\phi_{01}^{-7}\frac{m_2}{2r_{12}}\hat{A_1}_{ij}\hat{A_1}^{ij}
\\&+
\left(1+\frac{m_2}{2r_2}\right)^{-7} \hat{A_2}_{ij}\hat{A_2}^{ij} - 7 \phi_{02}^{-7} u_1^2\hat{A_2}_{ij}\hat{A_2}^{ij}\\& -7\phi_{02}^{-7}\frac{m_1}{2r_{12}}\hat{A_2}_{ij}\hat{A_2}^{ij}
\\&+\mathcal{O}\left(\frac{P^4}{r_{12}}\right)
+\mathcal{O}\left(\frac{J^4}{r_{12}}\right)+\mathcal{O}\left(\frac{J^2P^2}{r_{12}}\right)\\&
+\mathcal{O}\left(J^6\right)+\mathcal{O}\left(P^6\right)+\mathcal{O}\left(P^5J\right)+\mathcal{O}\left(J^5P\right)\\&
+\mathcal{O}\left(J^4P^2\right)+\mathcal{O}\left(J^3P^3\right)+\mathcal{O}\left(J^2P^4\right).
\end{split}
\end{align}
Eq.(\ref{Source term 2 BHs}) implies that at second order we have to solve the partial differential equation
\begin{align}
\begin{split}
\Delta u_2 =& +\frac{7}{8} \phi_{01}^{-7} u_1^1\hat{A_1}_{ij}\hat{A_1}^{ij} +\frac{7}{8}\phi_{01}^{-7}\frac{m_2}{2r_{12}}\hat{A_1}_{ij}\hat{A_1}^{ij}
\\&+\frac{7}{8} \phi_{02}^{-7} u_1^2\hat{A_2}_{ij}\hat{A_2}^{ij} +\frac{7}{8}\phi_{02}^{-7}\frac{m_1}{2r_{12}}\hat{A_2}_{ij}\hat{A_2}^{ij}.
\end{split}
\end{align}

Then if we write $u_2 = u_2^1 + u_2^2$ we can solve independently for the two BHs
\begin{align}
\begin{split}
\Delta u_2^1 =& +\frac{7}{8} \phi_{01}^{-7} u_1^1\hat{A_1}_{ij}\hat{A_1}^{ij} +\frac{7}{8}\phi_{01}^{-7}\frac{m_2}{2r_{12}}\hat{A_1}_{ij}\hat{A_1}^{ij}
\\\Delta u_2^2=&+\frac{7}{8} \phi_{02}^{-7} u_1^2\hat{A_2}_{ij}\hat{A_2}^{ij} +\frac{7}{8}\phi_{02}^{-7}\frac{m_1}{2r_{12}}\hat{A_2}_{ij}\hat{A_2}^{ij}.
\end{split}
\end{align}

And again, by performing a decomposition in spherical harmonics 
we can reduce this system of partial differential equations to a set of
independent ordinary differential equations.
We also point out that this method can be straightforwardly generalized to
an arbitrary number of BHs as follows
\begin{equation}
\Delta u_2^\alpha = +\frac{7}{8} \phi_{0\alpha}^{-7} u_1^\alpha\hat{A_\alpha}_{ij}\hat{A_\alpha}^{ij} +\sum_{\beta \neq \alpha}^{N_{BHs}}\frac{7}{8}\phi_{0\alpha}^{-7}\frac{m_\beta}{2r_{\alpha\beta}}\hat{A_\alpha}_{ij}\hat{A_\alpha}^{ij}.
\end{equation}
with $\alpha = 1,2,3,...,N_{BHs}$.

The total ADM mass at the second perturbative order is

\begin{align}\label{eq:MADM}
\begin{split}
M_{ADM}&=\sum_i^{N_{BHs}} \sum_{j\neq i}^{N_{BHs}}
\\&\frac{P_i^2}{2640000 m_i^5} \Big[(160 J_i^2 (1448 {R_{PJ}^{31}}^2+1448 {R_{PJ}^{32}}^2\\&
+1529{R_{PJ}^{33}}^2 -8000)\\&
+\frac{1650000 m_i^4 (s_{ij}-3 m_j)}{s_{ij}}-520157 m_i^2 P_i^2)\Big]
\\&-\frac{266 J_i^4}{825
   M_i^7}-\frac{2 J_i^2 (2 m_j-s_{ij})}{5 m_i^3 s_{ij}}-\frac{3 \norm{\textbf{P}_i\times \textbf{J}_i} }{110m_i^5}+m_i
\end{split}
\end{align}

where $R_{PJ}$ is the rotation matrix from the system with $\hat{P}=(0,0,1)$ to the system with $\hat{J}=(0,0,1)$ defined in Appendix \ref{Rotation_Matrix}.

The total ADM linear momentum and
angular momentum of the Bowen-York data are given by:
\begin{eqnarray}\label{eq:PJADM}
\vec P_{\rm ADM} &=& \sum_i^{N_{BHs}} \vec P_i,\\
\vec J_{\rm ADM} &=& \sum_i^{N_{BHs}} (\vec J_i + \vec r_i \times \vec P_i).
\end{eqnarray}

\section{Full Numerical Techniques}\label{sec:FN}

In order to perform the full numerical 
simulations we use the LazEv code\cite{Zlochower:2005bj} with 8th
order spatial finite differences \cite{Lousto:2007rj}, 4th order
Runge-Kutta time integration with a Courant factor $(dt/dx=1/4)$.

To compute the numerical initial data, we use the puncture
approach~\cite{Brandt97b} along with the {\sc  TwoPunctures}
~\cite{Ansorg:2004ds} code.
We use {\sc AHFinderDirect}~\cite{Thornburg2003:AH-finding} to locate
apparent horizons.  We measure the magnitude of the horizon spin 
$S_H$, using the {\it isolated horizon} algorithm 
as  implemented in Ref.~\cite{Campanelli:2006fy}.
We can then calculate the horizon
mass via the Christodoulou formula 
${m^H} = \sqrt{m_{\rm irr}^2 + S_H^2/(4 m_{\rm irr}^2)}\,,$
where $m_{\rm irr} = \sqrt{A_H/(16 \pi)}$ and $A_H$ is the surface area
of the horizon.

The {\sc Carpet}~\cite{Schnetter-etal-03b} mesh refinement driver
provides a ``moving boxes'' style of mesh refinement. In this
approach, refined grids of fixed size are arranged about the
coordinate centers of the holes.  The code then moves these fine
grids about the computational domain by following the trajectories of
the black holes.

The grid structure of our mesh refinements have a size of the largest
box for all simulations of $\pm400M$.  The number of points between 0
and 400 on the coarsest grid is XXX in nXXX (i.e. n100 has 100
points).  So, the grid spacing on the coarsest level is 400/XXX.  The
resolution in the wavezone is $100M/$XXX (i.e. n100 has $M/1.00$, n120
has $M/1.2$ and n144 has $M/1.44$) and the rest of the levels is
adjusted globally. For instance, the grid around one of the black holes
($m_1$) is fixed at $\pm0.6M$ in size and is the 9th refinement level.
Therefore the grid spacing is 400/XXX/$2^8$.

We evaluate eccentricity during evolution via the simple formula,
as a function of the separation of the holes, $d$,
$e_d=d^2\ddot{d}/m$, as given in \cite{Campanelli:2008nk}.

We also use the proper distance between the two horizons as measured
along the coordinate line joining the two punctures
\cite{Alcubierre:2004hr}, which we call the simple proper distance, or
$d_{\rm{spd}}$, below (note that the minimal geodesic does not
necessarily follow this line).

The extraction of gravitational radiation from the numerical
relativity simulations is performed using the formulas (22) and (23)
from \cite{Campanelli:1998jv} for the energy and linear momentum
radiated, respectively, in terms of the extracted Weyl scalar $\Psi_4$
at the observer location $R_{obs}=113M$. For angular momentum radiated
we use the formulas in \cite{Lousto:2007mh}.


\begin{table}
\caption{ 
  Initial data parameters for the base binary (2BH0)
  and the two coplanar (3BH1, 3BH2) configurations
with a third black hole at a distance $D$ from the binary along the
$x$-axis. $(x_i,y_i,z_i)$ and $(p^x_i,p^y_i,p_i^z)$ are the initial
position and momentum of the puncture
$i$, $m^p_i$ is the puncture mass parameter, $m^H_i$ is the horizon
mass, $M \Omega$ is the binary's orbital frequency, $d$ is the binary's initial coordinate separation
and $d_{\rm{spd}}$ is the binary's simple proper distance. Parameters not specified are zero.
}
\begin{ruledtabular}
\begin{tabular}{lccc}
Config     &  2BH0      &  3BH1      &  3BH2      \\
\hline
$x_1/M$ & -9.95027835 & -9.95027835 & -9.98428541 \\
$y_1/M$ & 3.96401481 & 3.96401481 & 3.96401481 \\
$p_1^x/M$ & -0.05706988 & -0.05705839 & -0.05705731 \\
$p_1^y/M$ & -0.00036813 & -0.02179566 & 0.02154356 \\
$m_1^p/M$ & 0.32546442 & 0.32362400 & 0.32359400 \\
$m_1^H/M$ & 0.33334615 & 0.33335960 & 0.33332705 \\
$x_2/M$ & -9.95027835 & -9.95027835 & -9.98428541 \\
$y_2/M$ & -3.96401481 & -3.96401481 & -3.96401481 \\
$p_2^x/M$ & 0.05706988 & 0.05708137 & 0.05708246 \\
$p_2^y/M$ & 0.00036813 & -0.02105940 & 0.02227982 \\
$m_2^p/M$ & 0.32546442 & 0.32362400 & 0.32362400 \\
$m_2^H/M$ & 0.33334654 & 0.33335185 & 0.33333734 \\
$d/M$ & 7.92802962 & 7.92802962 & 7.92802962 \\
$d_{\rm{spd}}/M$ & 10.55538506 & 10.65527971 & 10.65538017 \\
$x_3/M$ & -- & 19.76412526 & 19.73192339 \\
$y_3/M$ & -- & 0.00000000 & 0.00000000 \\
$p_3^x/M$ & -- & -0.00002299 & -0.00002515 \\
$p_3^y/M$ & -- & 0.04285506 & -0.04382339 \\
$m_3^p/M$ & -- & 0.32908500 & 0.32901500 \\
$m_3^H/M$ & -- & 0.33334994 & 0.33330884 \\
$M\Omega$ & 0.03273404 & 0.00586017 & 0.00590334 \\
$D/M$ & -- & 29.7144036 & 29.7162088 \\
\end{tabular}
\end{ruledtabular}
\label{table:ID-1}
\end{table}


\subsection{Two Black Holes Test}\label{sec:2BHTest}

In order to evaluate quantitatively the improvements of this next to
leading parameters $(\vec{P}_i,\vec{S}_i,1/d_i)$ expansion with
respect to the leading (labeled for the sake of simplicity second and
first order respectively), we compare the evolution of a binary black
hole system from initial data generated by these two expansions and
that of the ``exact'' TwoPunctures \cite{Ansorg:2004ds} numerical solver.

We will consider an equal mass, nonspinning binary with a separation
of the holes $d=12m$, where $m$ is the sum of the horizon masses, that
in preparation to use this binary in the three black holes case (3BH)
(See Fig.~\ref{fig:configs}), we will take as $m^H_i=1/3$. The orbital
parameters are taken as those of a quasicircular orbit
\cite{Healy:2017zqj} and are given in the first column of
Table~\ref{table:ID-1}, and labeled as 2BH0.

\begin{figure}
\begin{center}
\includegraphics[width=\columnwidth]{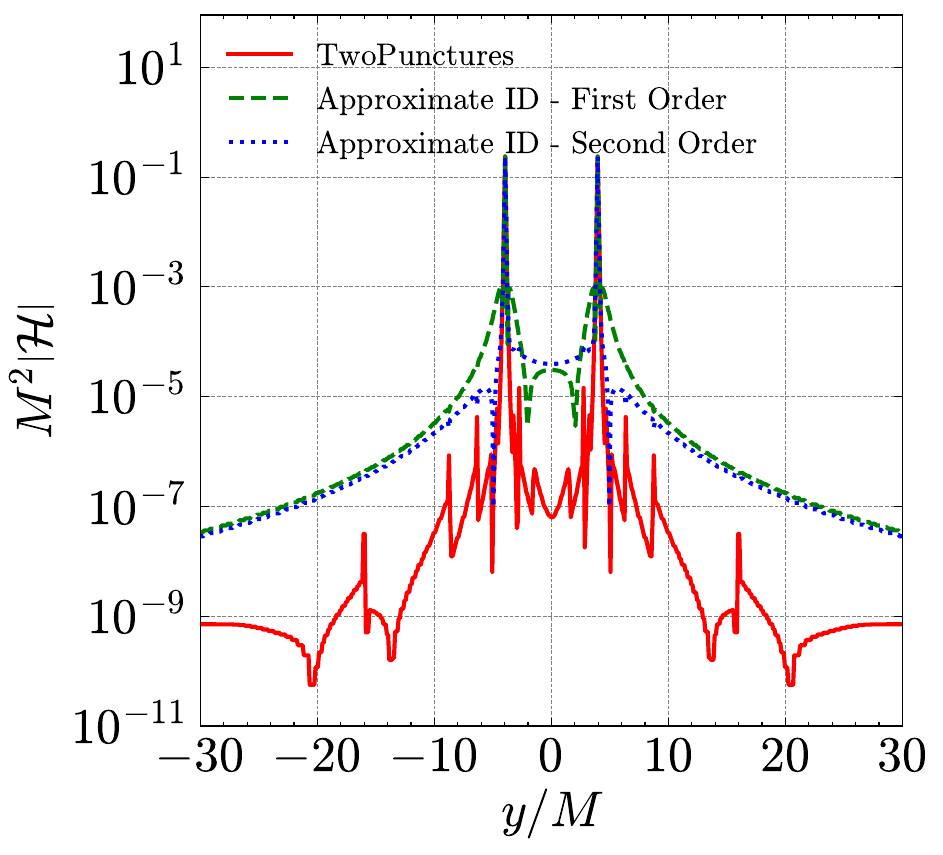}
  \caption{Violation of the constraints for the different sets of
  initial data considered here for the 2BH0 reference binary.}
\label{fig:Hct0}
\end{center}
\end{figure}

We first observe that placing those sets of initial data on the
numerical grid that will serve for its evolution, allow us to evaluate
the violations of the Hamiltonian constraint
$|\cal{H}|$. Figure~\ref{fig:Hct0} displays those violations along the
line joining the black holes. The spikes (in log-scale) shown
particularly in the TwoPunctures solution have to do with crossing the
zero-value at those points and the plotting of the Hamiltonian
magnitude $|\cal{H}|$. The first and second order approximation fall
well above the ``exact'' solution, with the second order improving
on the first order violations around the black holes and asymptotically
away.

\begin{figure}
\begin{center}
\includegraphics[width=\columnwidth]{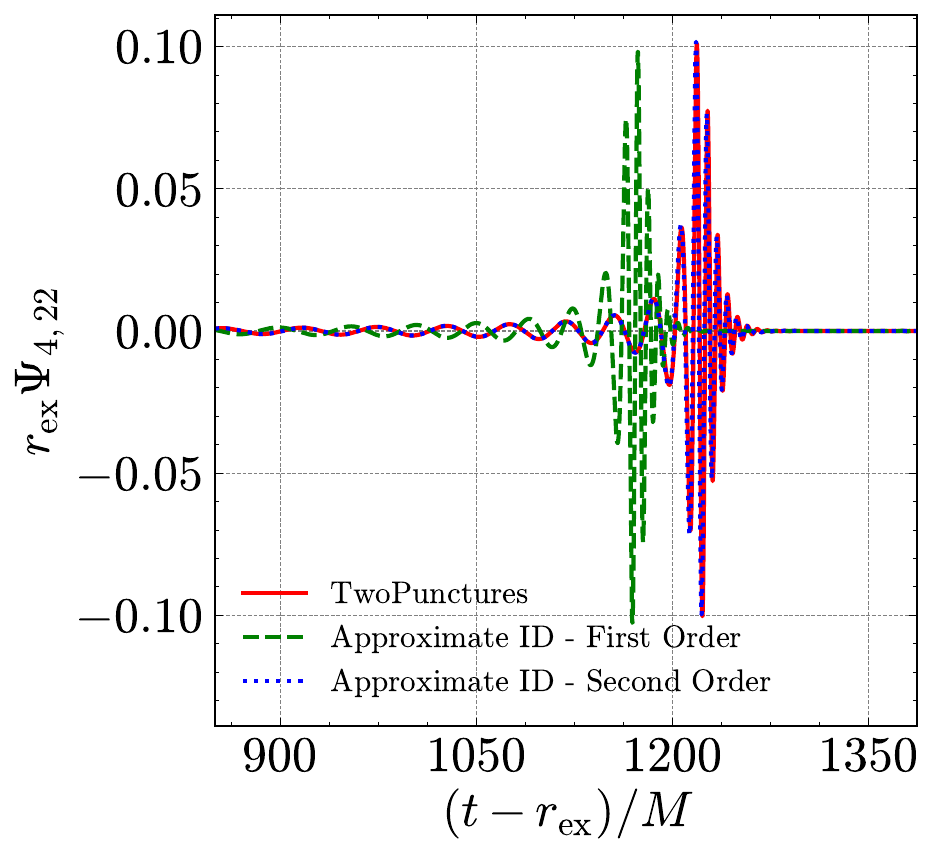}
\caption{Weyl scalar $\psi_4$ extracted at $r_{ex}=113M$ from the
  evolution of the 
    binary system started with the three different sets of initial data.}
\label{fig:psi42bh}
\end{center}
\end{figure}

\begin{figure}
\begin{center}
\includegraphics[width=\columnwidth]{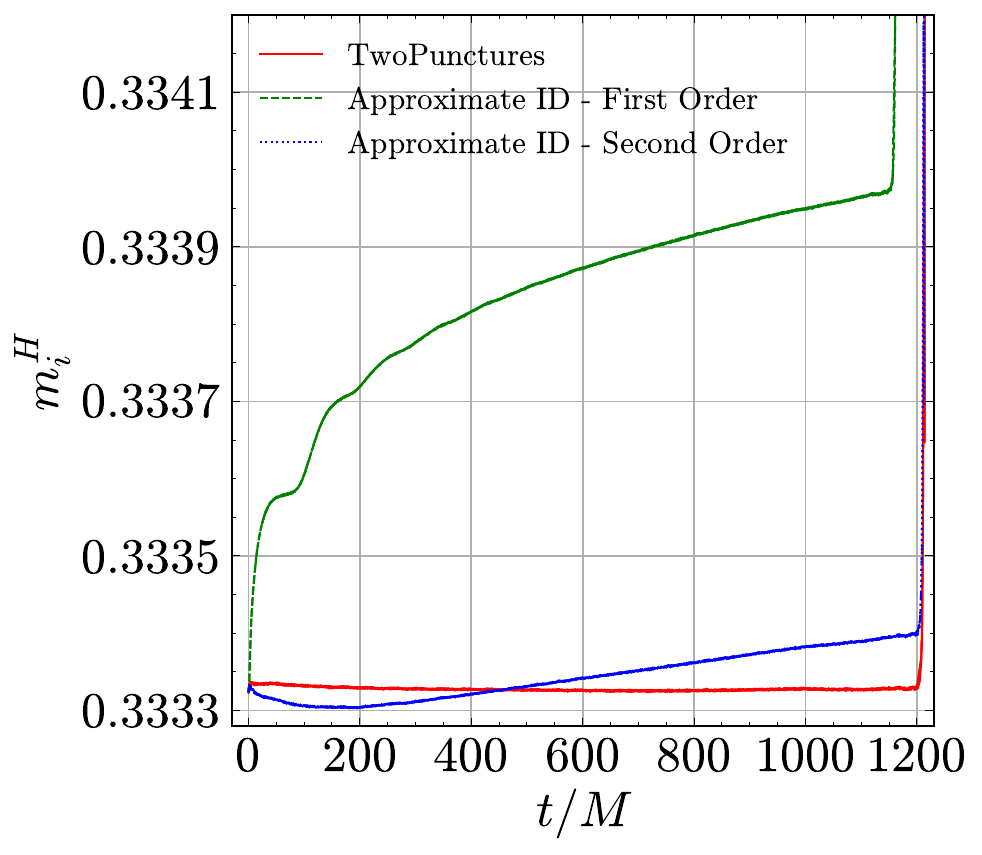}
  \caption{Evolution of the black holes horizon masses starting from the
    same normalization. Differences are due to different initial
    violations of the constraints for the sets of approximate
  initial data considered here.}
\label{fig:mh2bh}
\end{center}
\end{figure}

The evolution of these initial data leads to potentially different
tracks and hence waveforms. A comparative of the three cases of
initial data considered here (but using the same binary parameters as
in Table~\ref{table:ID-1}, 2BH0) is given in Fig.~\ref{fig:psi42bh}
where we observe the close match of the second order and ``exact''
TwoPunctures data in comparison with the first order case. This later
difference (already observed in Fig. 1 of Ref.~\cite{Lousto:2007rj})
can be in part traced back to the effects of the violations of the
Hamiltonian constraint in the initial data that propagates in the
numerical grid or is accreted by the black holes.  In fact we can
observe this effect in the evolution of the individual horizon masses
until merger in Fig.~\ref{fig:mh2bh}. That would lead to deviations in
their relative tracks explaining the differences in their
corresponding waveforms in Fig.~\ref{fig:psi42bh}.


We supplement the information of the initial data here with
another measure of the initial data quality as is the computation of
the ``binding'' energy of the two black holes $E_b=M_{ADM}-m$ as the
difference of the total ADM mass $M_{ADM}$ and the sum of the horizon masses
$m=m_1^H+m_2^H$.  We compare here its computation via the
TwoPuncture numerical solution to the Hamiltonian constraint to
the first and second order analytic approximations as given in
Eq.~(\ref{eq:MADM}). For our binary separated by $d=12m$ 
we find $E_b^N=-0.00588611$ for the TwoPuncture numerical solution
while $E_b^f=-0.00350055$, and $E_b^s=-0.00486506$, for the first and
second order solutions, representing a 40\% and 17\% differences,
respectively.

\section{Three black holes evolutions}\label{sec:3BH}

\begin{figure*}
\begin{center}
\includegraphics[width=1.5\columnwidth]{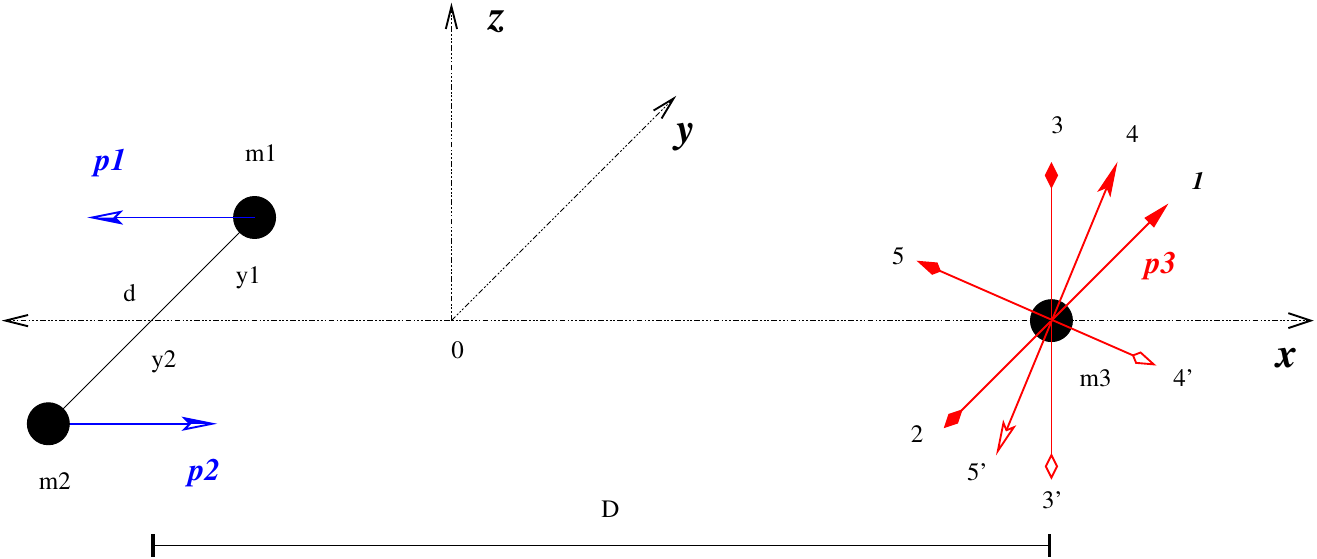}
  \caption{Initial configurations considered
    for the three black hole evolutions, labeled as 3BH1-5
  (3', 4', 5' are quasi-symmetric counterparts).}
\label{fig:configs}
\end{center}
\end{figure*}

We will consider a series of prototypical simulations involving three
black holes. In this first exploration we will consider a hierarchical
system with the inner binary at an initial separation of $12m=8M$ and a
third black hole at separation $30M$. All black holes in this first
set will initially have equal masses (as measured by their individual
horizons) and no spins, but with different relative orbital orientations.
This set up is depicted in Fig.~\ref{fig:configs}.

%

As a first estimate of the orbital periods we can use the Keplerian
expression $P=2\pi/\Omega$ where the orbital frequency is 
$\Omega=m/r^{3/2}$. Thus for the binary (at $r=12m=8M)$, we find an
initial period of $P_B=174M$ while for the orbit of the third black
hole (at $r=30M)$, a period of $P_3=1032M$. 
From the quasicircular initial orbit \cite{Healy:2017zqj} from the
third post-Newtonian order (3PN) given in the Table~\ref{table:ID-1}
we find $P_B^{3PN}=192M$ and $P_3^{3PN}=1072M$, respectively.  What we
measure from the simulation tracks is in close correspondence with
those values, i.e. $P_B^{NR}\approx205M$ and $P_3^{NR}\approx1060M$.

In order to chose parameters leading to small initial eccentricities
we first consider the inner binary as isolated and apply the
quasicircular formulas of Ref.~\cite{Healy:2017zqj} to obtain the
parameters reported in the first column of Table~\ref{table:ID-1} and
referred to as 2BH0. Once we have the inner binary parameters we
apply the same quasicircular criteria to 
the outer orbit of the third black hole with an effective spinning
black hole having the added masses and angular momentum of the
inner binary.
In practice this process works to provide low enough eccentricities
$(e\lesssim0.05)$ for our initial study purposes.

\subsection{Three black holes in a hierarchical system}\label{sec:3BHp}

To start exploring this vast parameter space we have chosen to
consider two coplanar cases, when the third black hole orbit is 
corotating with the binary (3BH1) and when it is counter-rotating
(3BH2). Those parameters are given in Table~\ref{table:ID-1}.
We also consider precessing cases with the third black hole
momentum perpendicular to the orbital plane of the binary (3BH3)
and at $\pm45$ degrees with respect to that (3BH4 and 3BH5), as
depicted in Fig.~\ref{fig:configs}. In all cases we considered
the quasicircular orbit of the third black hole with the 
inner binary as an effective single black hole.
The corresponding parameters for these cases
are given in Table~\ref{table:ID-4}.


\begin{table}
\caption{Initial data parameters for the precessing three black hole
cases (3BH3, 3BH4 and 3BH5).}
\begin{ruledtabular}
\begin{tabular}{lccc}
Config     &  3BH3       &  3BH4      &  3BH5      \\
\hline
$x_1/M$ & -9.96709434 & -9.95516516 & -9.97921106 \\
$y_1/M$ & 3.96401481 & 3.96401481 & 3.96401481 \\
$z_1/M$ & 0.00000000 & 0.00000000 & 0.00000000 \\
$p_1^x/M$ & -0.05705789 & -0.05705825 & -0.05705749 \\
$p_1^y/M$ & -0.00036813 & -0.01556934 & 0.01507516 \\
$p_1^z/M$ & -0.02166824 & -0.01520121 & 0.01544329 \\
$m_1^p/M$ & 0.32359400 & 0.32359400 & 0.32359400 \\
$m_1^H/M$ & 0.33332740 & 0.33332841 & 0.33332691 \\
$x_2/M$ & -9.96709434 & -9.95516516 & -9.97921106 \\
$y_2/M$ & -3.96401481 & -3.96401481 & -3.96401481 \\
$z_2/M$ & 0.00000000 & 0.00000000 & 0.00000000 \\
$p_2^x/M$ & 0.05708188 & 0.05708152 & 0.05708228 \\
$p_2^y/M$ & 0.00036813 & -0.01483308 & 0.01581142 \\
$p_2^z/M$ & -0.02166824 & -0.01520121 & 0.01544329 \\
$m_2^p/M$ & 0.32359400 & 0.32359400 & 0.32359400 \\
$m_2^H/M$ & 0.33332869 & 0.33332239 & 0.33333438 \\
$d/M$ & 7.92802962 & 7.92802962 & 7.92802962 \\
$d_{\rm{spd}}/M$ & 10.65491937 & 10.65499784 & 10.65513690 \\
$x_3/M$ & 19.74820210 & 19.75949795 & 19.73672848 \\
$y_3/M$ & 0.00000000 & 0.00000000 & 0.00000000 \\
$z_3/M$ & 0.00000000 & 0.00000000 & 0.00000000 \\
$p_3^x/M$ & -0.00002398 & -0.00002326 & -0.00002479 \\
$p_3^y/M$ & 0.00000000 & 0.03040241 & -0.03088657 \\
$p_3^z/M$ & 0.04333649 & 0.03040241 & -0.03088657 \\
$m_3^p/M$ & 0.32910500 & 0.32906500 & 0.32906500 \\
$m_3^H/M$ & 0.33338457 & 0.33333346 & 0.33335498 \\
$M\Omega$ & 0.00588143 & 0.00586633 & 0.00589685 \\
$D/M$ & 29.7152964 & 29.7146631 & 29.7159395 \\
\end{tabular}
\end{ruledtabular}
\label{table:ID-4}
\end{table}

\begin{figure}
\begin{center}
\includegraphics[width=\columnwidth]{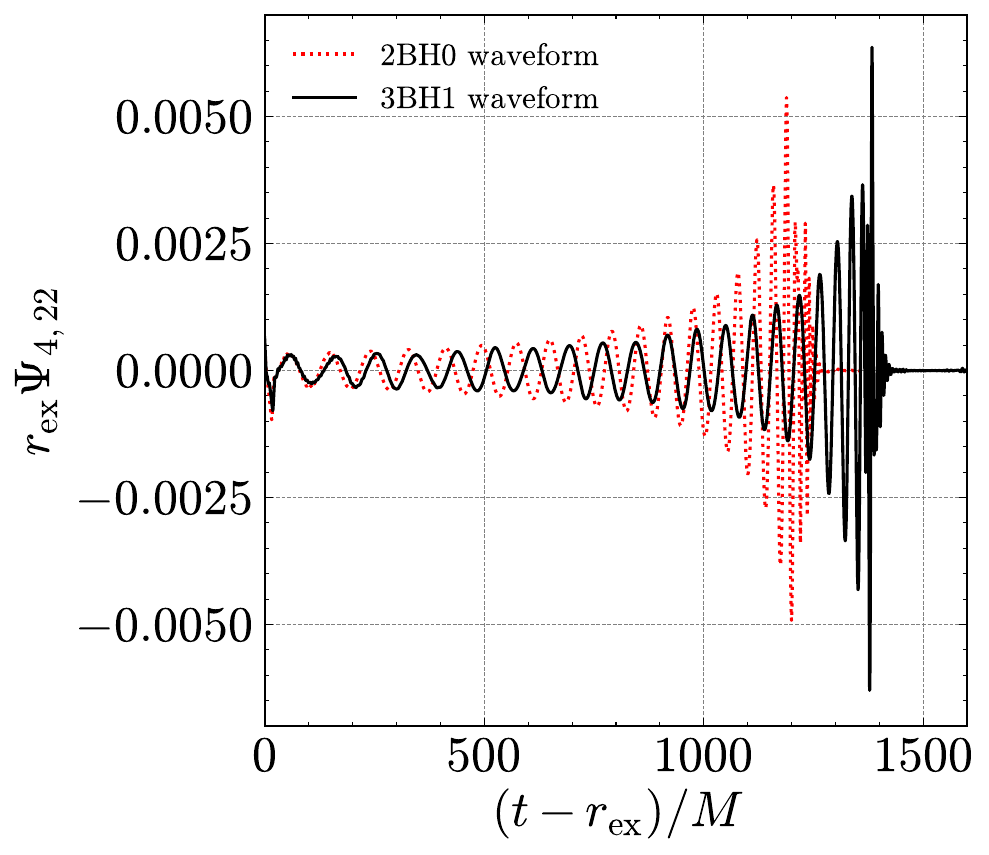}
\caption{Waveforms generated by the case 3BH1 in
  comparison with the isolated binary 2BH0.}
\label{fig:ipw}
\end{center}
\end{figure}

\begin{figure}
  \begin{center}
  \includegraphics[width=0.85\columnwidth]{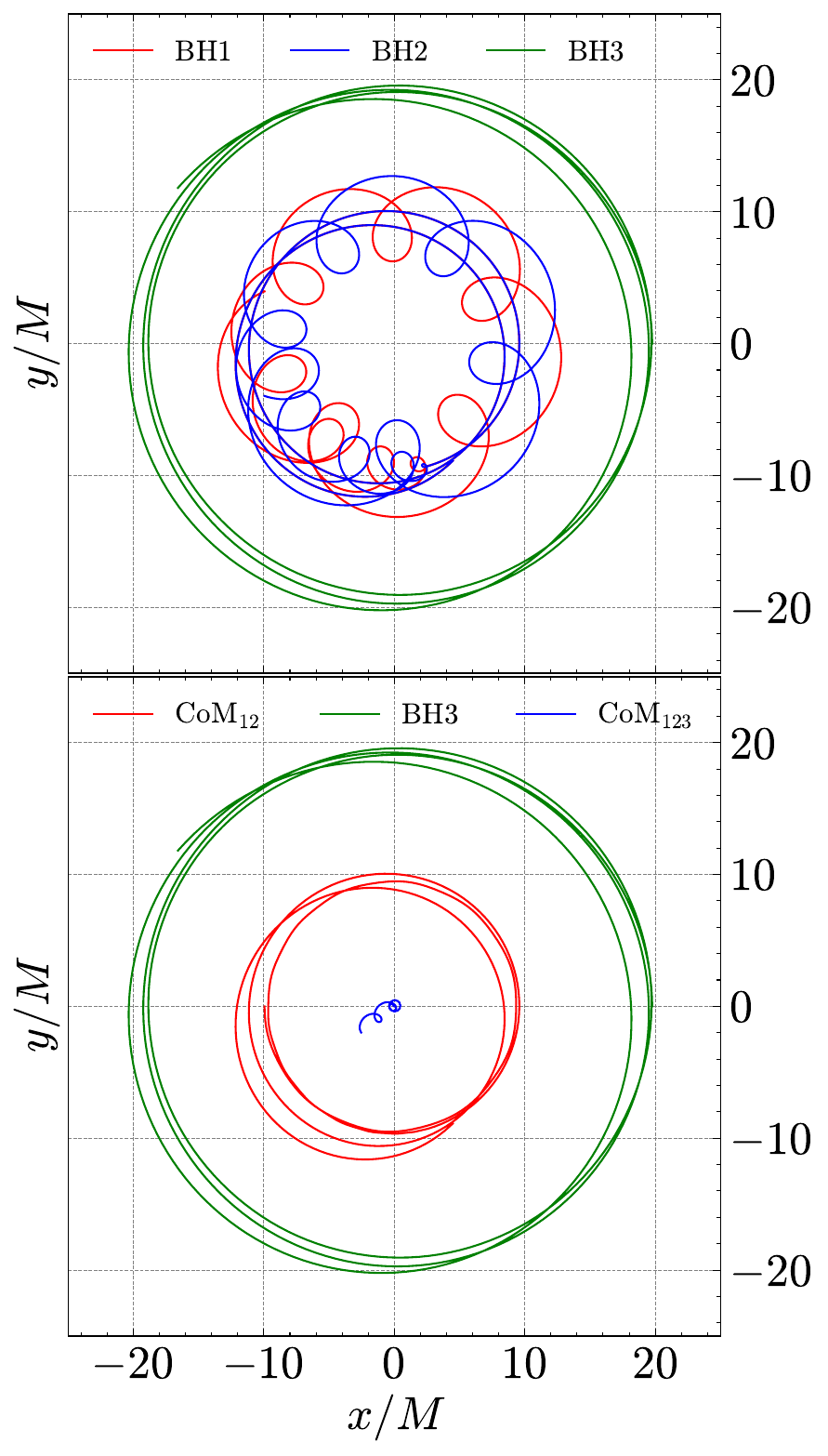}
  \caption{Trajectories of the coplanar case 3BH1 and the evolution of
    the center of masses of the binary and of the three black holes.}
\label{fig:ip}
\end{center}
\end{figure}

In Fig.~\ref{fig:ipw} we display the extracted waveform of
the three black hole simulation 3BH1. The gravitational
radiation is completely dominated by the inner binary. The
difference with an isolated binary is given by the delay
in the merger due to the presence of the third black hole.
Similar results are obtained for the 3BH2-5 cases.
Another effect is the motion of the binary and its
merger product around the
center of mass of the triple system, as displayed in
Fig.~\ref{fig:ip}. This leads to a mixing of modes
as seen by a fixed observer location, but its effects
can be disentangled with techniques like those used
in Refs.~\cite{Woodford:2019tlo,Healy:2020vre}.
%

%

\begin{figure}
\begin{center}
\includegraphics[width=\columnwidth]{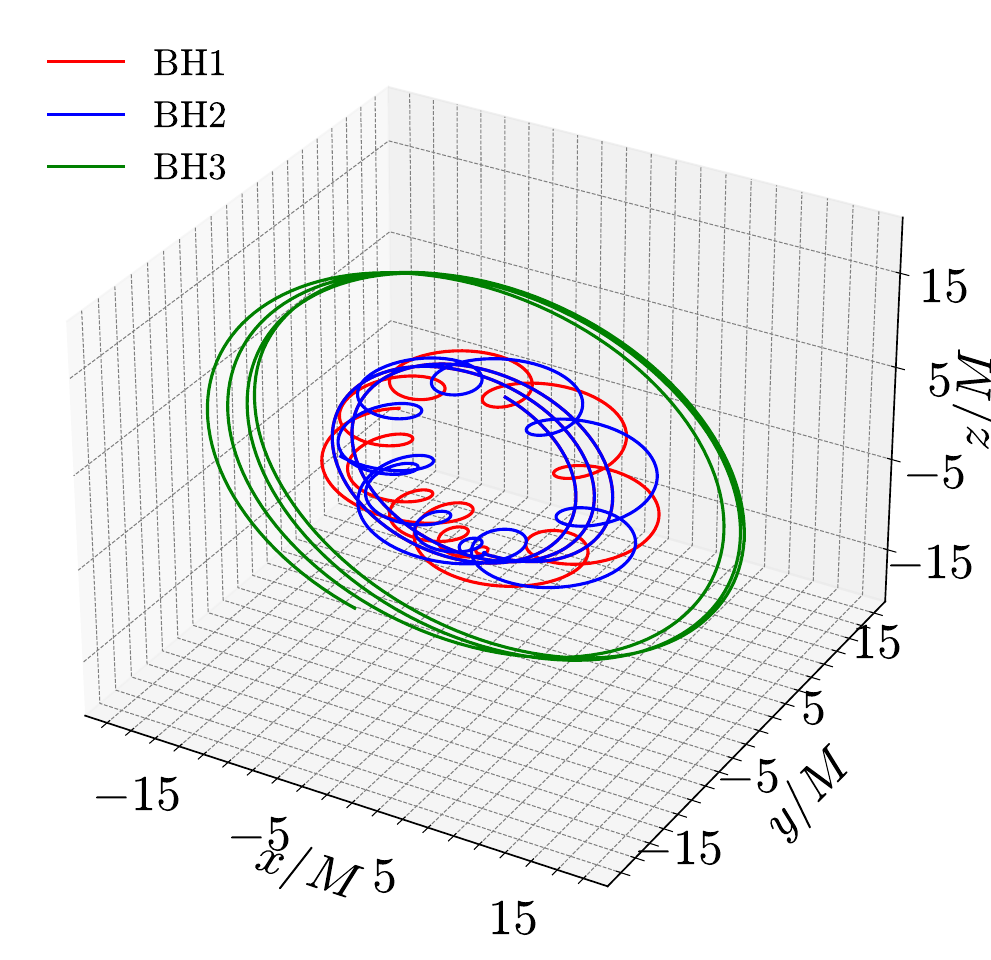}
  \caption{Trajectories of the fully precessing case 3BH3.}
\label{fig:iiip}
\end{center}
\end{figure}

In Fig.~\ref{fig:iiip} we also display the trajectories of
the three black holes in the fully precessing
case 3BH3 in three dimensions. They clearly display the precession
of the third black hole orbital plane over the three orbits of our
simulation.
%

In Table~\ref{tab:3BHp} we report the merger times of the
first five cases studied here. We first note the clear delay
of the merger of 3BH1-5 with respect to the isolated binary
2BH0. We then note the relatively weak dependence of the
merger times and number of orbits on the
orientation of the orbit, at this initial separation of
the third hole, $D=30M$.

\begin{table}
\caption{Number of orbits to merger and merger time of the inner binary for different orbital orientations of the third black hole. Cases 3BH1-5}
\begin{ruledtabular}
\begin{tabular}{lll}
label & $\#$orbits & $t_{\text{merger}}/M$ \\
\hline
2BH0 & 9.949  & 1216.9\\
3BH1 & 10.637 & 1376.6 \\
3BH2 & 10.821 & 1419.9 \\
3BH3 & 10.523 & 1341.2 \\
3BH4 & 10.582 & 1358.0 \\
3BH5 & 10.705 & 1387.5 \\
\end{tabular}
\end{ruledtabular}
\label{tab:3BHp}
\end{table}

\begin{figure}
\begin{center}
  \includegraphics[width=\columnwidth]{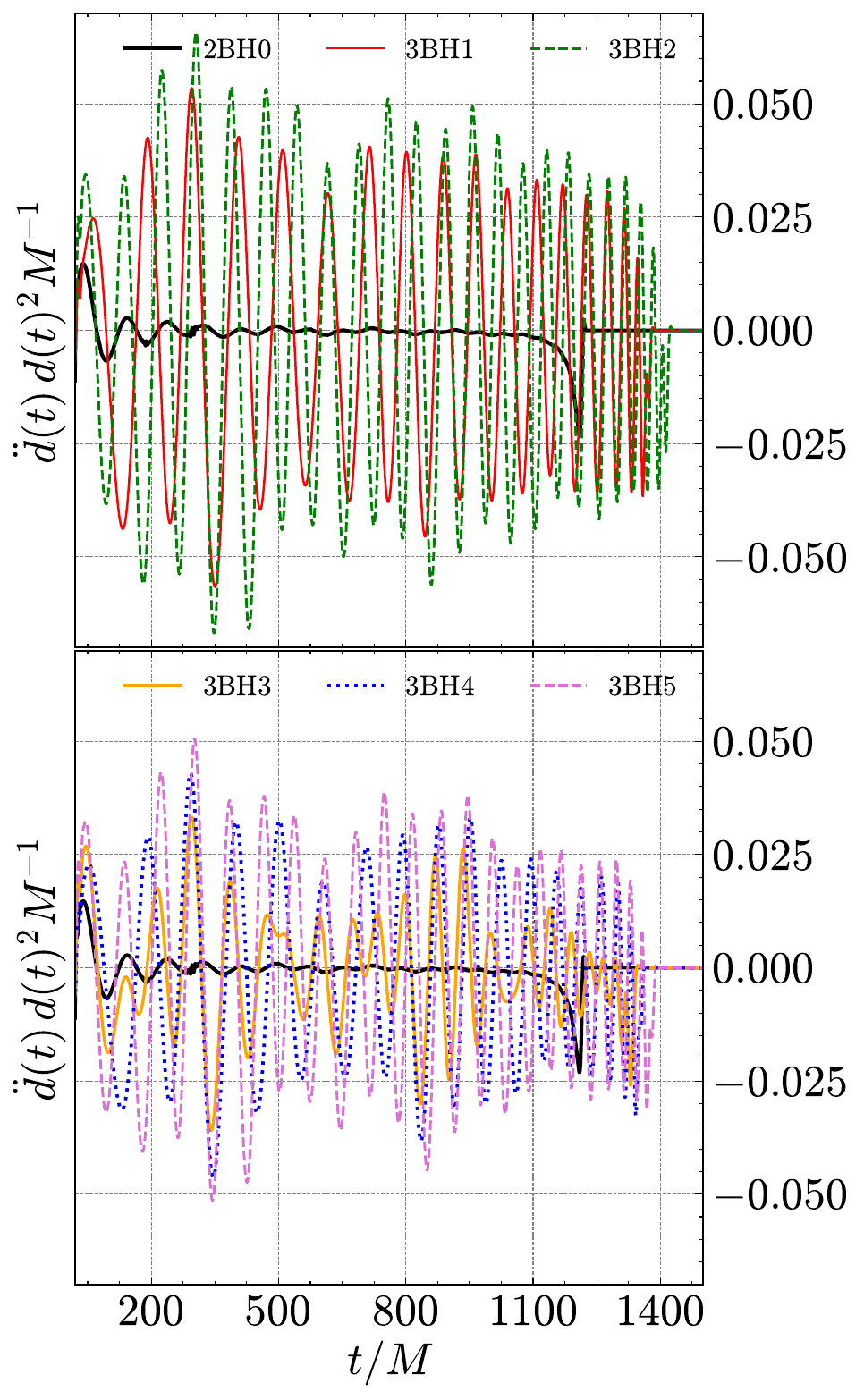}
\caption{Eccentricity evolution of the inner binary
  as measured by $d^2\ddot{d}(t)$
  for the triple black hole cases 3BH1-5 and the isolated reference
  binary 2BH0.}
\label{fig:B1ecc}
\end{center}
\end{figure}

The other interesting property that we want to study here is
the evolution of the eccentricity of the binary due to the
presence of the third black hole in a hierarchical orbit
around the binary. In Fig.~\ref{fig:B1ecc}
we display the instantaneous \cite{Campanelli:2008nk}
eccentricity $e(t)\approx e\cos(\Omega t)\approx d^2\ddot{d}/M$,
of the inner binary for the three black hole cases 3BH1-5
and the isolated reference binary 2BH0 (B1).
We first observe that the amplitude of the eccentricity slightly
decrease notably during evolution and presents a modulation
with the third black hole orbital frequency.

\begin{figure}
\begin{center}
  \includegraphics[width=\columnwidth]{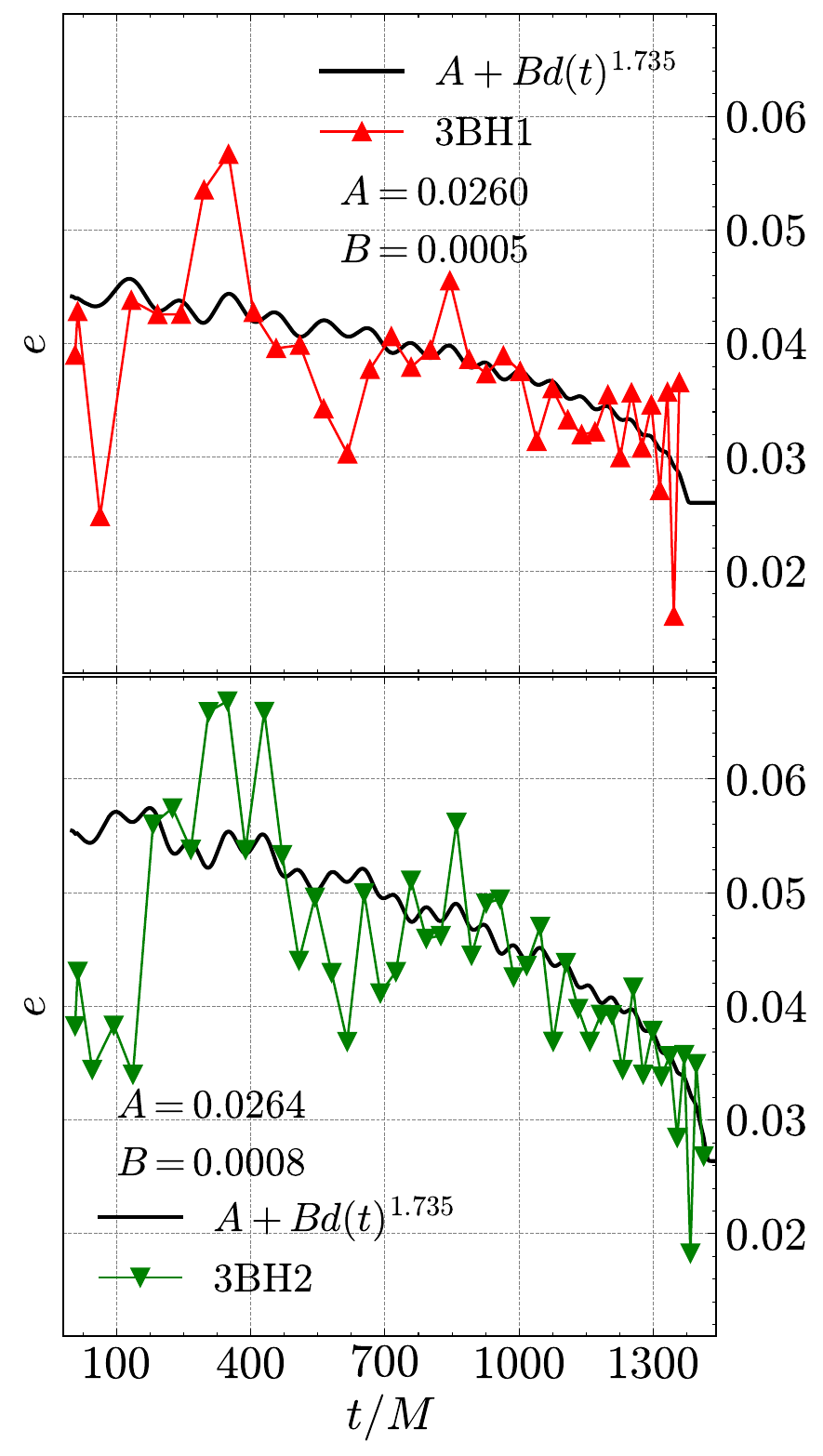}
\caption{Eccentricity evolution of the inner binary
  as measured by the amplitude of $d^2\ddot{d}(t)$
  for the triple black hole cases 3BH1-2 and the
  expected decay $d^{1.735}$ of Ref.~\cite{Lousto:2015uwa}.
  (Note the zoom factor $\times10$ on the right panel). 
\label{fig:evst}}
\end{center}
\end{figure}

In order to visualize better the evolution of the eccentricity we take
the values of the extremes of oscillations per orbit to model the
$e(t)\approx e\cos(\Omega t)$ dependence and extract the values of $e$
per each half orbit of the coplanar cases 3BH1 (corotating orbits) and
3BH2 (counterrotating orbits).  The results of this analysis are
displayed in Fig.~\ref{fig:evst}.

This is first contrasted with what
we expect from an isolated binary on the grounds of the
decay of the eccentricity with
the instantaneous separation as $\sim d^{1.735}$, found from
numerical simulations, see also Fig.~9 in Ref.~\cite{Lousto:2015uwa}.
We observe that even if the inner binary
starts at a relatively close separation, $12m$, leading to ten orbits
before merger compared to the nearly fifty orbits of the simulation
analyzed in Ref.~\cite{Lousto:2015uwa}, a general trend towards
decrease can be observed.  Particularly closer to merger, during the
last few orbits, we see a decrease in the eccentricity in
as expected on the fact that at those close
separations the relative influence of the third black hole should be
reduced.
We have also verified that the 1PN predictions
\cite{Peters:1964} that should show a decay of the eccentricity with
the instantaneous separation as $\sim d^{19/12}$, give very close
results to those displayed in Fig.~\ref{fig:evst}. See also recent
2PN studies in Ref.~\cite{Datta:2023uln}.

Another feature that appears in both cases displayed in Fig.~\ref{fig:evst}
is a modulation superposed over an steady decrease of the eccentricity.
This modulation has a period of around $\sim500M$ which seems to correspond
to the semi-periods of the third black hole, that we estimated above to
be initially of the order of $\approx1060M$. It also bears resemblance
to a beating frequency of the two orbital motions
$(\Omega_{2BH}-\Omega_{3BH})/2\approx468M$.

\begin{figure}
\begin{center}
  \includegraphics[width=\columnwidth]{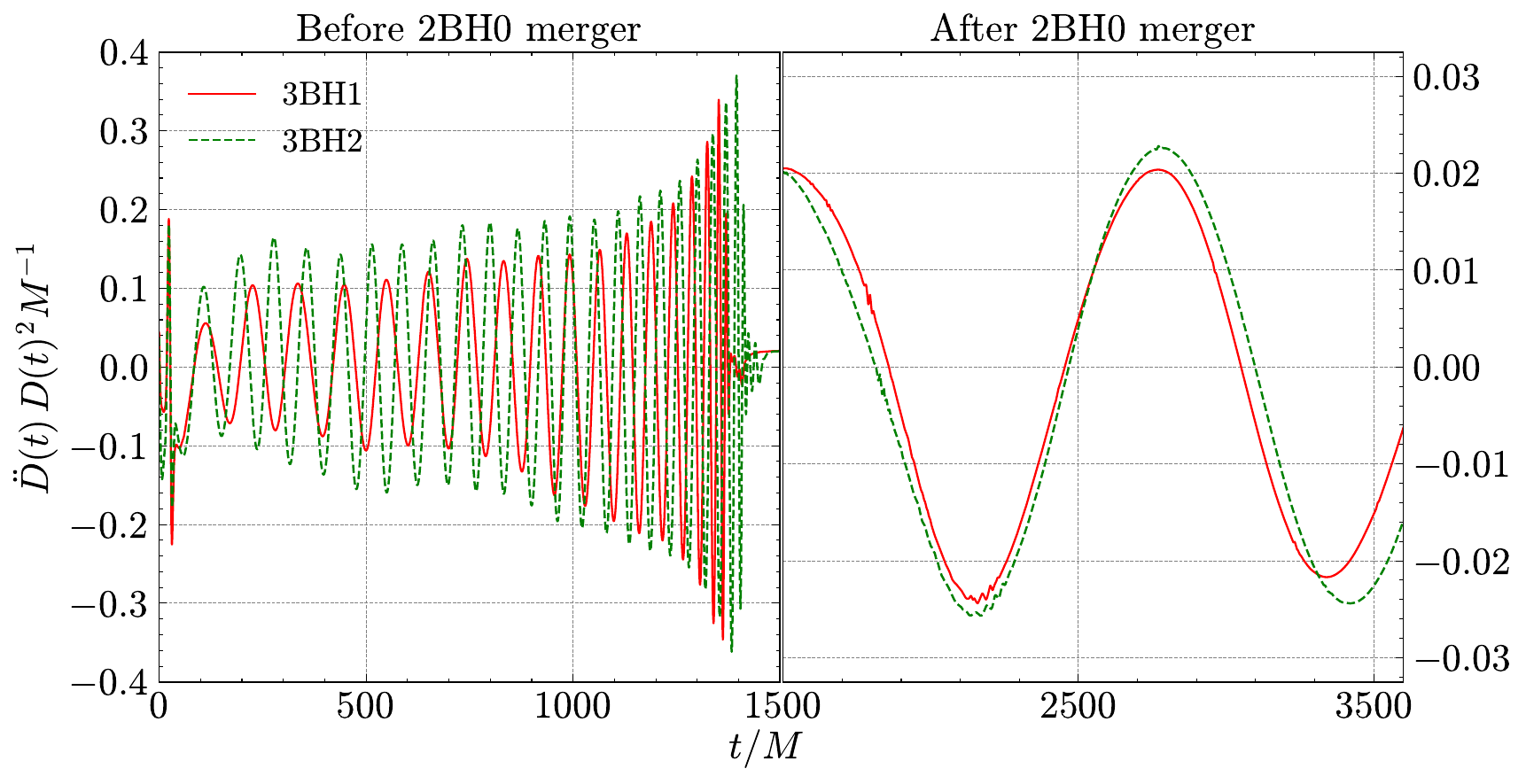}
\caption{Eccentricity evolution of the outer black hole
  as measured by the amplitude of $D^2\ddot{D}(t)$
  for the three coplanar black hole cases 3BH1-2.
\label{fig:3BHe}}
\end{center}
\end{figure}

Finally, we can look at the eccentricity evolution of the orbit of
the third black hole. Before merger we can refer its 
motion around the
center of mass of the binary system, as displayed in the 
bottom of Fig.~\ref{fig:ip} and then after the merger of the
the inner binary to its remnant, as displayed on the right panel of
Fig.~\ref{fig:3BHe}.
We note that the eccentricity measure from the center of mass of the binary
seems to grow in time and reaches relatively large values before merger.
This seems to be an effect of the use of the coordinates of the
center of mass as a reference of this
extended system. We note that right after merger the eccentricity
measure produces an order of magnitude less eccentricity for the
subsequent two orbits and with values more in line with what we
expect and found for the inner binary studies above.
Qualitatively similar results have been found for the precessing
cases 3BH3-5.


\subsection{Numerical convergence}\label{sec:convergence}

Here we explore the dependence of the previous results on the
numerical resolution of the finite difference integrations to
perform the evolutions of three black holes. To that end
we perform a series of three simulation of the representative
case 3BH1, with increasing global resolutions by factors of
1.2, namely the original simulation at n100 resolution and
two additional ones at n120 and n144 resolutions. The results
of such simulations is summarized in Table~\ref{tab:convergence},
were we report the merger times and number of orbits of the
inner binary as defined by its trajectories approaching at
a distance of $d_m=0.7M$ (this corresponds closely to the first
appearance of a common apparent horizon within a $\Delta t\sim 5M$,
as we verified directly for 3BH1).

\begin{table}
\label{tab:convergence}
\caption{Convergence of number of orbits and merger time for the 3BH1 configuration using three resolutions. Richardson extrapolation is used to determine convergence order and infinitely extrapolated values. We point out that the difference between different resolutions is smaller than differences between the different configurations.}
\begin{tabular}{lccc}
\toprule
resolution & $\#$orbits & $t_{\rm{merger}}/M$ \\
\hline
n100 & 10.637 & 1376.6 \\
n120 & 10.603 & 1370.4 \\
n144 & 10.597 & 1369.6 \\
Inf. Extrap. & 10.596 & 1369.5 \\
\hline
Inf.$-$ n100 & -0.041 & -7.1 \\
$\%$ difference & -0.387 & -0.518\\
\hline
Conv. order & 9.51 & 11.23 \\
\hline
\end{tabular}
\end{table}

We observe that those values align in a convergence order leading
to high powers of convergence, as computed by the formulas
(5a)-(5c) of \cite{Lousto:2019lyf}, still comparable to the
expected 8th order convergence from the spatial finite difference
stencils used in our integration algorithm. 
The relevant point here is that the differences of the n100
simulations values we use as a basis to extract conclusions about
delays in merger times and number of orbits to merger to
its (Richardson's) extrapolation to infinite resolution is
very small compared to the physical changes we observe. We hence
conclude they are a numerically reliable result and will keep
using this n100 resolution as the standard for the following
studies.


\subsection{The Distance dependence to the third black hole}\label{sec:3BHd}

Given the weak sensitivity of the binary evolution with the direction
of the third black hole momentum, we will next explore how the merger
times and eccentricity evolution of the binary vary versus the initial
separation of the outer black hole. For that end we look again for
quasicircular effective parameters at different initial separations
as given in Table~\ref{table:ID-7}.

\begin{table*}
\caption{Initial data parameters for coplanar-corotating configurations with a third black hole placed at different distances $D$ from the binary along the $x$-axis, 3BHD1-5.}
\begin{ruledtabular}
\begin{tabular}{lccccc}
Config & 3BHD1 & 3BHD2 & 3BHD3 & 3BHD4 & 3BHD5 \\
\hline
$x_1/M$ & -8.28225635 & -11.61795003 & -13.28540069 & -14.95270161 & -19.95406346 \\
$y_1/M$ & 3.96401481 & 3.96401481 & 3.96401481 & 3.96401481 & 3.96401481 \\
$p_1^x/M$ & -0.05705000 & -0.05706264 & -0.05706503 & -0.05706647 & -0.05706844 \\
$p_1^y/M$ & -0.02408677 & -0.02005564 & -0.01867731 & -0.01755115 & -0.01511026 \\
$m_1^p/M$ & 0.32325400 & 0.32386400 & 0.32406400 & 0.32421600 & 0.32453600 \\
$m_1^H/M$ & 0.33335247 & 0.33333709 & 0.33334068 & 0.33333886 & 0.33335088 \\
$x_2/M$ & -8.28225635 & -11.61795003 & -13.28540069 & -14.95270161 & -19.95406346 \\
$y_2/M$ & -3.96401481 & -3.96401481 & -3.96401481 & -3.96401481 & -3.96401481 \\
$p_2^x/M$ & 0.05708976 & 0.05707712 & 0.05707474 & 0.05707329 & 0.05707133 \\
$p_2^y/M$ & -0.02335051 & -0.01931938 & -0.01794106 & -0.01681489 & -0.01437400 \\
$m_2^p/M$ & 0.32325400 & 0.32386400 & 0.32406400 & 0.32421600 & 0.32453600 \\
$m_2^H/M$ & 0.33334355 & 0.33332930 & 0.33333270 & 0.33333175 & 0.33334414 \\
$d/M$ & 7.92802962 & 7.92802962 & 7.92802962 & 7.92802962 & 7.92802962 \\
$d_{\rm{spd}}/M$ & 10.67520743 & 10.64064151 & 10.63015429 & 10.62193283 & 10.60525218 \\
$x_3/M$ & 16.46307522 & 23.06578078 & 26.36782091 & 29.67012192 & 39.57796287 \\
$y_3/M$ & 0.00000000 & 0.00000000 & 0.00000000 & 0.00000000 & 0.00000000 \\
$p_3^x/M$ & -0.00003976 & -0.00001448 & -0.00000970 & -0.00000682 & -0.00000288 \\
$p_3^y/M$ & 0.04743728 & 0.03937502 & 0.03661837 & 0.03436605 & 0.02948426 \\
$m_3^p/M$ & 0.32823500 & 0.32968500 & 0.33014500 & 0.33048500 & 0.33116543 \\
$m_3^H/M$ & 0.33333394 & 0.33334810 & 0.33335404 & 0.33333988 & 0.33330813 \\
$M\Omega$ & 0.00763429 & 0.00465268 & 0.00384912 & 0.00323775 & 0.00211854 \\
$D/M$ & 24.7453316 & 34.6837309 & 39.6532216 & 44.6228235 & 59.5320263 \\
\end{tabular}
\end{ruledtabular}
\label{table:ID-7}
\end{table*}


We are interested in studying the effect the third hole has on
the inner binary dynamics. In particular how it affects the
merger, if prompts or delays it. In Table~\ref{tab:3BHd} we
give the results of our simulations versus the initial third
black hole distance to the binary's center of mass. We find
a clear trend towards the delay of the merger, in both measures,
the merger time and the number of orbits as measured by the
tracks of the holes and using a definition of merger when the
binary distance reaches $d=0.7M$ (which corresponds closely
to the formation of a common horizon).  
\begin{table}
\caption{Number of orbits to merger and merger time of the inner binary for different initial separation of the third black hole. Cases 3BHD1-5}
\begin{ruledtabular}
\begin{tabular}{lll}
  $D/M$ & $\#$orbits & $t_{\text{merger}}/M$ \\
\hline
 30 & 10.63  & 1376.6 \\
 35 & 10.40  & 1323.6 \\
 40 & 10.25  & 1292.6 \\
 45 & 10.18  & 1275.0 \\
 60 & 10.09  & 1250.5  \\
 $\infty$ & 9.94915  & 1216.875\\
\end{tabular}
\end{ruledtabular}
\label{tab:3BHd}
\end{table}

%
\begin{figure}
  \begin{center}
      \includegraphics[width=0.9\columnwidth]{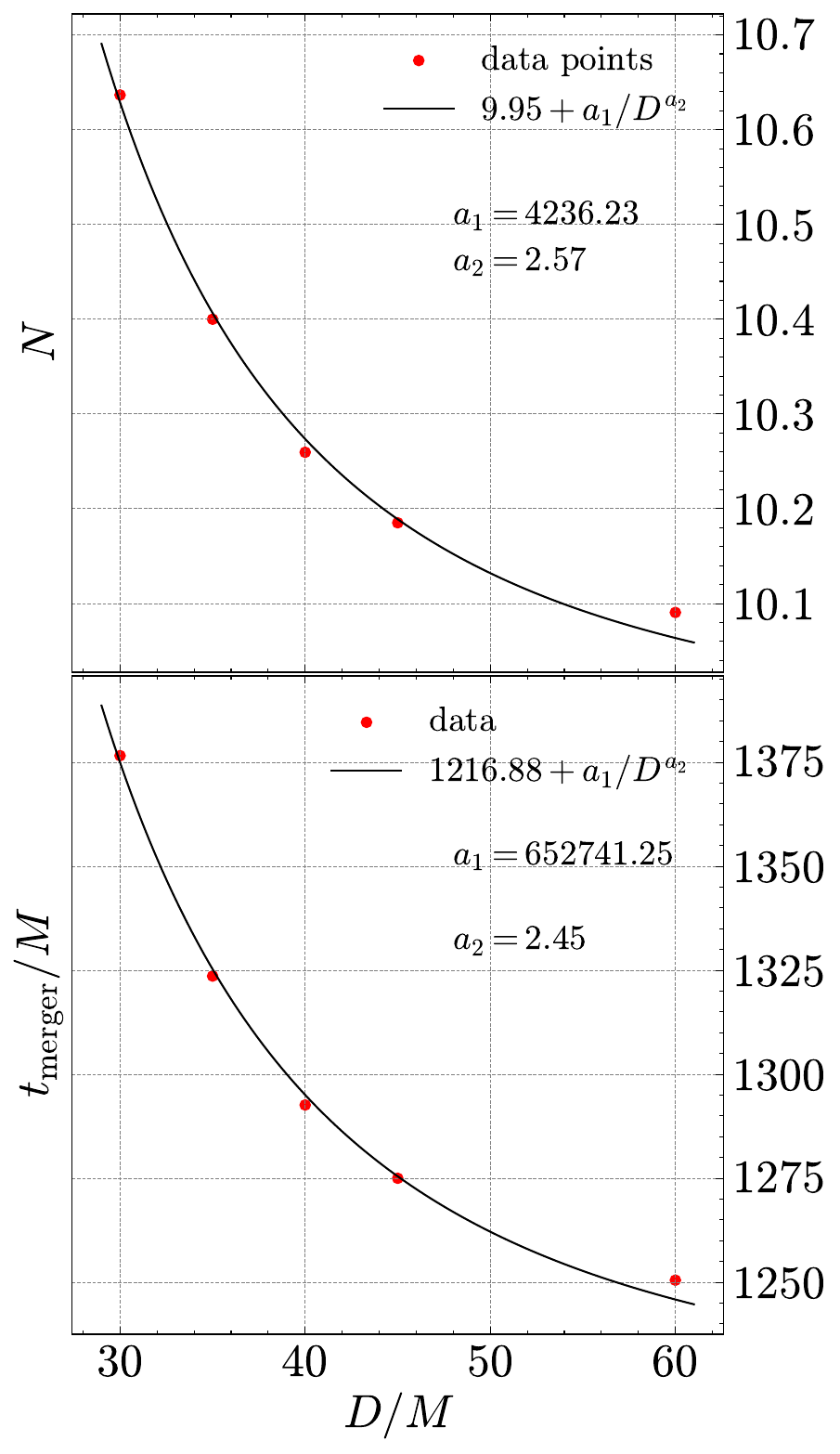}
  \caption{Fit to a functional dependence $2BH0+a1/D^{a2}$.}
\label{fig:3BHd}
\end{center}
\end{figure}

In order to model the merger delay as a function of the initial
distance to the third black hole we consider deviations with
respect to the merger time and number of orbits to merger
isolated binary, 2BH0. We thus fit a dependence to the data
in Table~\ref{tab:3BHd} of the form $2BH0+a1/D^{a2}$. The
results are displayed in Fig.~\ref{fig:3BHd} and lead to
a consistent dependence of the form $1/D^{2.5}$.
\begin{figure}[h!]
\begin{center}
  \includegraphics[width=\columnwidth]{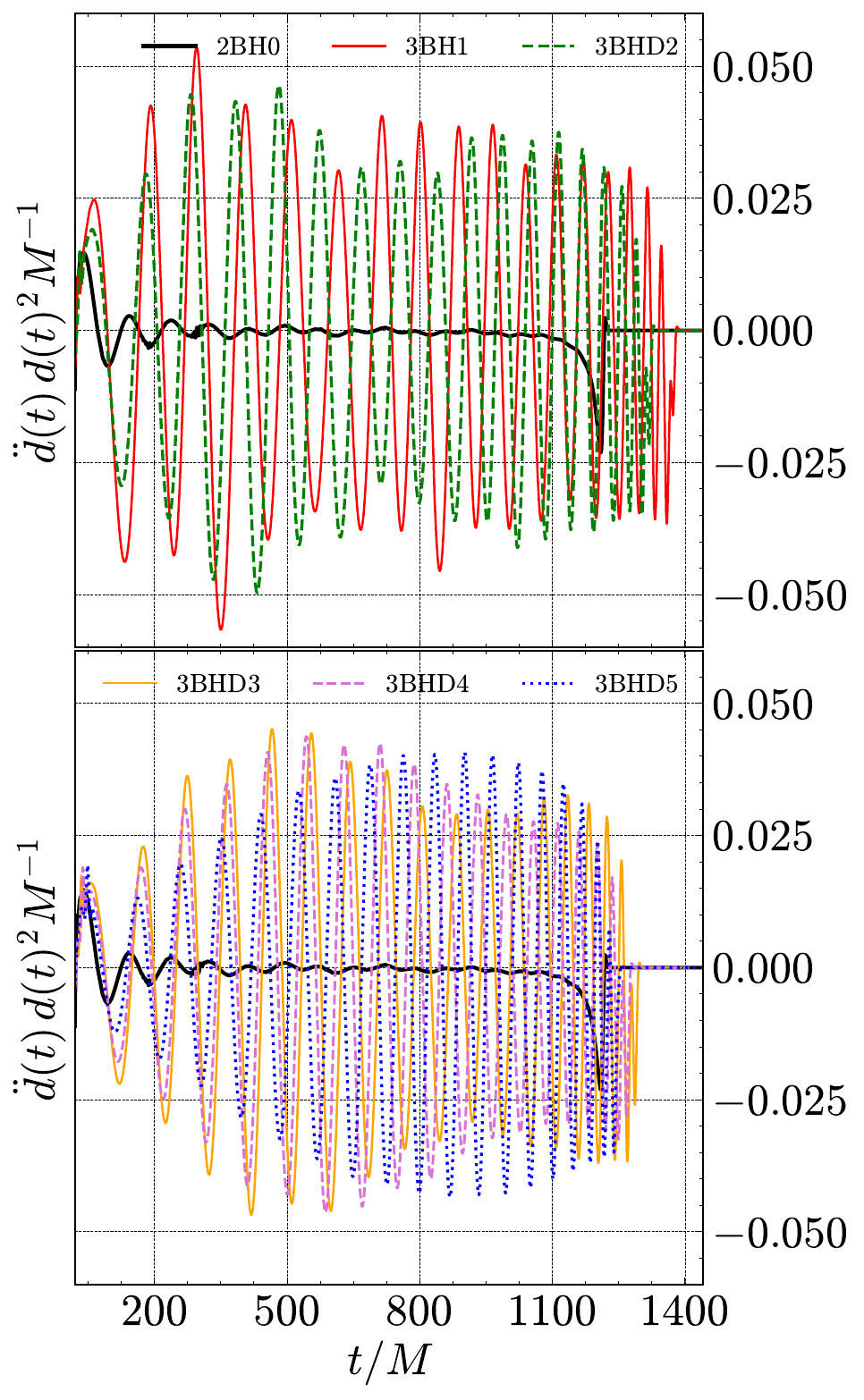}
\caption{Eccentricity evolution of the inner binary
  as measured by $d^2\ddot{d}(t)$
  for the triple black hole cases 3BH1, 3BHD2-5 and the isolated reference
  binary 2BH0.}
\label{fig:B1eccd}
\end{center}
\end{figure}

We again study the instantaneous eccentricity evolution of the
inner binary as we vary the orbital distance of the third black
hole. The results are displayed in Fig.~\ref{fig:B1eccd}.
While the initial magnitude of the eccentricity is
due to the choice of the orbital parameters their evolution
shows a trend towards reduction for all cases, particularly
very close to merger.

%

\section{Conclusions and Discussion}\label{sec:Discussion}

Although full numerical solutions to three black holes initial data
have been presented in
Refs.~\cite{Galaviz:2010mx,Bai:2011za,Imbrogno:2021xrh} we found a
valid and practical option, validated for the two black hole cases,
to provide analytic initial data for prompt
use and with enough accuracy for current exploratory studies.

We next revisited the triple black hole scenario to study their
merging times and eccentricity evolution.  We found that the third
black hole delays the merger of the binary by an amount inversely
proportional to a power of the distance, $\sim1/D^{2.5}$. This
behavior was not clearly observed in some of the configurations
simulated in a previous work \cite{Lousto:2007rj}, due to the
closeness of the cases studied that lead to a prompt breakdown of the
binary, as we also observe here if we start the third black hole
closer to $\approx30M$.  We also note here that the $\sim1/D^{2.5}$
dependence can be associated to a 5th post-Newtonian correction and
its leading tidal effects on the inner binary
waveforms~\cite{Flanagan:2007ix}.

A delay in the merger time of the binary due to the presence of the
third black hole has also been observed in the Post-Newtonian
approximation \cite{Galaviz:2010te} considering much larger
separations of the binary $(130M-170M)$ and to the third black hole up
to $(10\,000M)$, thus representing a complementary study to the one
presented here.

The presence of a nearby third black hole also seems to confirm a
decay of any residual inner binary eccentricity and to induce a subtle
modulation with about a half the period of the third black hole orbit
around the binary. Note that in Ref.~\cite{Naoz:2012bx} it was studied
with post-Newtonian techniques~\cite{Lousto:2007ji} resonant
eccentricity excitation in hierarchical three-body systems, another
complementary study to that presented here.

The next natural exploration of 3BH interactions with our formalism
involves the inclusion of spins in the inner binary, the unequal mass
ratio to consider binaries in the field of a much larger black hole,
and the scattering effects of a passing third black hole. Those will
be covered in a forthcoming study.

\begin{acknowledgments}
The authors thank James Healy, Hiroyuki Nakano, and Yosef Zlochower
for useful discussions. The authors also gratefully acknowledge
the National Science Foundation (NSF) for financial support from Grant
No.\ PHY-1912632 and PHY-2207920.  Computational resources were also
provided by the New Horizons, Blue Sky, Green Prairies, and White
Lagoon clusters at the CCRG-Rochester Institute of Technology, which
were supported by NSF grants No.\ PHY-0722703, No.\ DMS-0820923,
No.\ AST-1028087, No.\ PHY-1229173, No.\ PHY-1726215, and
No.\ PHY-2018420.  This work used the ACCESS allocation TG-PHY060027N,
founded by NSF, and project PHY20007 Frontera, an NSF-funded Petascale
computing system at the Texas Advanced Computing Center (TACC).
\end{acknowledgments}
\appendix

\section{Rotation matrix}\label{Rotation_Matrix}

When we want to study the perturbation of the Hamiltonian constraint at second order we encounter terms of interaction between the momentum $\textbf{P}$ and spin $\textbf{J}$ of the same black hole. Since this perturbative solution requires to choose a specific axis with respect to which we write the spherical harmonics, we need to be able to write the angle between the position $\textbf{r}$ and, for example, the spin in terms of the angular coordinates taken starting from $\textbf{P}$ as $z$ axis.

In order to do so let's consider the momentum versor $\hat{\textbf{P}}$ and spin versor $\hat{\textbf{J}}$ in a certain coordinate system

\begin{equation}
\hat{\textbf{P}} =
\begin{bmatrix}
P_x \\  
P_y \\  
P_z   
\end{bmatrix} \ \ \ \ \ \ \ \ 
\hat{\textbf{J}} =
\begin{bmatrix}
J_x \\  
J_y \\  
J_z   
\end{bmatrix}
\end{equation}

Let's call the matrix $R_{G\textbf{P}}$ the matrix that rotates the vector $\hat{\textbf{P}}$ into the vector $\hat{\textbf{z}}$
with 
\begin{equation}
\hat{\textbf{z}} =
\begin{bmatrix}
0 \\  
0 \\  
1   
\end{bmatrix}
\end{equation}

This matrix is constructed through

\begin{equation}
R_{GP} = I + V + \frac{V\cdot V}{1+C}
\end{equation}

where
\begin{equation}
V=
\begin{bmatrix}
0 & -V_z & V_y \\  
V_z & 0 & -V_x \\  
-V_y & V_x & 0   
\end{bmatrix} 
\end{equation}

and 

\begin{equation}
\textbf{V}=\hat{\textbf{P}}\times\hat{\textbf{z}} \ \ \ \ \ \ \ \ C=\hat{\textbf{P}}\cdot\hat{\textbf{z}}
\end{equation}

This is also the matrix that transforms the coordinates of a given vector in the reference system G to the one in which the z axis is aligned along $\hat{\textbf{P}}$ (which we call P).

Analogously we find $R_{G\textbf{J}}$ and $R_{G\textbf{J}\times\textbf{P}}$.

Once we have these matrices we can combine them to find $R_{\textbf{P}\textbf{J}}$, $R_{\textbf{P}\textbf{J}\times\textbf{P}}$ and $R_{\textbf{J}\textbf{J}\times\textbf{P}}$.

Now let's consider for example the unit vector $\hat{\textbf{n}}$ which has coordinates 

\begin{equation}
\hat{\textbf{n}} =
\begin{bmatrix}
n_{xP} \\  
n_{yP} \\  
n_{zP}   
\end{bmatrix}
\end{equation}

in the P system.

Then the coordinates of $\hat{\textbf{n}}$ in the J system are 

\begin{equation}
\begin{bmatrix}
n_{xJ} \\  
n_{yJ} \\  
n_{zJ}   
\end{bmatrix}= R_{\textbf{P}\textbf{J}}
\begin{bmatrix}
n_{xP} \\  
n_{yP} \\  
n_{zP}  
\end{bmatrix}
\end{equation}

In particular we are only interested in the 3rd coordinate $n_{zJ}$ which is

\begin{eqnarray}
  n_{zJ} &&= R_{\textbf{P}\textbf{J}}^{31}n_{xP} + R_{\textbf{P}\textbf{J}}^{32}n_{yP} + R_{\textbf{P}\textbf{J}}^{33}n_{zP} \\
  &&=\sin{\theta_\textbf{P}}(R_{\textbf{P}\textbf{J}}^{31}\cos{\phi_\textbf{P}} + R_{\textbf{P}\textbf{J}}^{32}\sin{\phi_\textbf{P}})+ R_{\textbf{P}\textbf{J}}^{33}\cos{\theta_\textbf{P}}
  \nonumber
\end{eqnarray}

Applying this procedure for all the cases we need we finally obtain

\begin{align}
\begin{split}
  n_{zJ} =& \sin{\theta_\textbf{P}}(R_{\textbf{P}\textbf{J}}^{31}\cos{\phi_\textbf{P}} + R_{\textbf{P}\textbf{J}}^{32}\sin{\phi_\textbf{P}})\\
  &+ R_{\textbf{P}\textbf{J}}^{33}\cos{\theta_\textbf{P}}\\
  n_{z\textbf{J}\times\textbf{P}} =& \sin{\theta_\textbf{P}}(R_{\textbf{P}\textbf{J}\times\textbf{P}}^{31}\cos{\phi_\textbf{P}} + R_{\textbf{P}\textbf{J}\times\textbf{P}}^{32}\sin{\phi_\textbf{P}})\\
  &+ R_{\textbf{P}\textbf{J}\times\textbf{P}}^{33}\cos{\theta_\textbf{P}}\\
  n_{z\textbf{J}\times\textbf{P}} =& \sin{\theta_\textbf{J}}(R_{\textbf{J}\textbf{J}\times\textbf{P}}^{31}\cos{\phi_\textbf{J}} + R_{\textbf{J}\textbf{J}\times\textbf{P}}^{32}\sin{\phi_\textbf{J}})\\
  &+ R_{\textbf{J}\textbf{J}\times\textbf{P}}^{33}\cos{\theta_\textbf{J}}
\end{split}
\end{align}

To determine let's say the angle $\phi_\textbf{P}$ (an analogous argument holds for $\phi_\textbf{J}$) we make use of the matrix $R_{G\textbf{P}}$ as follows. Let's say that in in the system of coordinates $G$ the coordinates of $\hat{\textbf{n}}$ are 

\begin{equation}
\hat{\textbf{n}} =
\begin{bmatrix}
n_{xG} \\  
n_{yG} \\  
n_{zG}   
\end{bmatrix}
\end{equation}

Then we have,

\begin{align}
\begin{split}
\tan{\phi_\textbf{P}} =& {\frac{\sin{\phi_\textbf{P}}}{\cos{\phi_\textbf{P}}}}
= \frac{(R_{G\textbf{P}}\cdot \textbf{n})_x}{(R_{G\textbf{P}}\cdot \textbf{n})_y}
\end{split}
\end{align}

From this we can obtain $\phi_\textbf{P}$ (and $\phi_\textbf{J}$).

\bibliographystyle{apsrev4-1}
\bibliography{../../../Bibtex/references.bib}

\begin{thebibliography}{33}%
\makeatletter
\providecommand \@ifxundefined [1]{%
 \@ifx{#1\undefined}
}%
\providecommand \@ifnum [1]{%
 \ifnum #1\expandafter \@firstoftwo
 \else \expandafter \@secondoftwo
 \fi
}%
\providecommand \@ifx [1]{%
 \ifx #1\expandafter \@firstoftwo
 \else \expandafter \@secondoftwo
 \fi
}%
\providecommand \natexlab [1]{#1}%
\providecommand \enquote  [1]{``#1''}%
\providecommand \bibnamefont  [1]{#1}%
\providecommand \bibfnamefont [1]{#1}%
\providecommand \citenamefont [1]{#1}%
\providecommand \href@noop [0]{\@secondoftwo}%
\providecommand \href [0]{\begingroup \@sanitize@url \@href}%
\providecommand \@href[1]{\@@startlink{#1}\@@href}%
\providecommand \@@href[1]{\endgroup#1\@@endlink}%
\providecommand \@sanitize@url [0]{\catcode `\\12\catcode `\$12\catcode
  `\&12\catcode `\#12\catcode `\^12\catcode `\_12\catcode `\%12\relax}%
\providecommand \@@startlink[1]{}%
\providecommand \@@endlink[0]{}%
\providecommand \url  [0]{\begingroup\@sanitize@url \@url }%
\providecommand \@url [1]{\endgroup\@href {#1}{\urlprefix }}%
\providecommand \urlprefix  [0]{URL }%
\providecommand \Eprint [0]{\href }%
\providecommand \doibase [0]{http://dx.doi.org/}%
\providecommand \selectlanguage [0]{\@gobble}%
\providecommand \bibinfo  [0]{\@secondoftwo}%
\providecommand \bibfield  [0]{\@secondoftwo}%
\providecommand \translation [1]{[#1]}%
\providecommand \BibitemOpen [0]{}%
\providecommand \bibitemStop [0]{}%
\providecommand \bibitemNoStop [0]{.\EOS\space}%
\providecommand \EOS [0]{\spacefactor3000\relax}%
\providecommand \BibitemShut  [1]{\csname bibitem#1\endcsname}%
\let\auto@bib@innerbib\@empty
\bibitem [{\citenamefont {Gayathri}\ \emph {et~al.}(2022)\citenamefont
  {Gayathri}, \citenamefont {Healy}, \citenamefont {Lange}, \citenamefont
  {O'Brien}, \citenamefont {Szczepanczyk}, \citenamefont {Bartos},
  \citenamefont {Campanelli}, \citenamefont {Klimenko}, \citenamefont
  {Lousto},\ and\ \citenamefont {O'Shaughnessy}}]{Gayathri:2020coq}%
  \BibitemOpen
  \bibfield  {author} {\bibinfo {author} {\bibfnamefont {V.}~\bibnamefont
  {Gayathri}}, \bibinfo {author} {\bibfnamefont {J.}~\bibnamefont {Healy}},
  \bibinfo {author} {\bibfnamefont {J.}~\bibnamefont {Lange}}, \bibinfo
  {author} {\bibfnamefont {B.}~\bibnamefont {O'Brien}}, \bibinfo {author}
  {\bibfnamefont {M.}~\bibnamefont {Szczepanczyk}}, \bibinfo {author}
  {\bibfnamefont {I.}~\bibnamefont {Bartos}}, \bibinfo {author} {\bibfnamefont
  {M.}~\bibnamefont {Campanelli}}, \bibinfo {author} {\bibfnamefont
  {S.}~\bibnamefont {Klimenko}}, \bibinfo {author} {\bibfnamefont {C.~O.}\
  \bibnamefont {Lousto}}, \ and\ \bibinfo {author} {\bibfnamefont
  {R.}~\bibnamefont {O'Shaughnessy}},\ }\href {\doibase
  10.1038/s41550-021-01568-w} {\bibfield  {journal} {\bibinfo  {journal}
  {Nature Astron.}\ }\textbf {\bibinfo {volume} {6}},\ \bibinfo {pages} {344}
  (\bibinfo {year} {2022})},\ \Eprint {http://arxiv.org/abs/2009.05461}
  {arXiv:2009.05461 [astro-ph.HE]} \BibitemShut {NoStop}%
\bibitem [{\citenamefont {Yu}\ \emph {et~al.}(2020)\citenamefont {Yu},
  \citenamefont {Ma}, \citenamefont {Giesler},\ and\ \citenamefont
  {Chen}}]{Yu:2020iqj}%
  \BibitemOpen
  \bibfield  {author} {\bibinfo {author} {\bibfnamefont {H.}~\bibnamefont
  {Yu}}, \bibinfo {author} {\bibfnamefont {S.}~\bibnamefont {Ma}}, \bibinfo
  {author} {\bibfnamefont {M.}~\bibnamefont {Giesler}}, \ and\ \bibinfo
  {author} {\bibfnamefont {Y.}~\bibnamefont {Chen}},\ }\href {\doibase
  10.1103/PhysRevD.102.123009} {\bibfield  {journal} {\bibinfo  {journal}
  {Phys. Rev. D}\ }\textbf {\bibinfo {volume} {102}},\ \bibinfo {pages}
  {123009} (\bibinfo {year} {2020})},\ \Eprint
  {http://arxiv.org/abs/2007.12978} {arXiv:2007.12978 [gr-qc]} \BibitemShut
  {NoStop}%
\bibitem [{\citenamefont {Dall'Amico}\ \emph {et~al.}(2021)\citenamefont
  {Dall'Amico}, \citenamefont {Mapelli}, \citenamefont {Di~Carlo},
  \citenamefont {Bouffanais}, \citenamefont {Rastello}, \citenamefont
  {Santoliquido}, \citenamefont {Ballone},\ and\ \citenamefont
  {Sedda}}]{DallAmico:2021umv}%
  \BibitemOpen
  \bibfield  {author} {\bibinfo {author} {\bibfnamefont {M.}~\bibnamefont
  {Dall'Amico}}, \bibinfo {author} {\bibfnamefont {M.}~\bibnamefont {Mapelli}},
  \bibinfo {author} {\bibfnamefont {U.~N.}\ \bibnamefont {Di~Carlo}}, \bibinfo
  {author} {\bibfnamefont {Y.}~\bibnamefont {Bouffanais}}, \bibinfo {author}
  {\bibfnamefont {S.}~\bibnamefont {Rastello}}, \bibinfo {author}
  {\bibfnamefont {F.}~\bibnamefont {Santoliquido}}, \bibinfo {author}
  {\bibfnamefont {A.}~\bibnamefont {Ballone}}, \ and\ \bibinfo {author}
  {\bibfnamefont {M.~A.}\ \bibnamefont {Sedda}},\ }\href {\doibase
  10.1093/mnras/stab2783} {\bibfield  {journal} {\bibinfo  {journal} {Mon. Not.
  Roy. Astron. Soc.}\ }\textbf {\bibinfo {volume} {508}},\ \bibinfo {pages}
  {3045} (\bibinfo {year} {2021})},\ \Eprint {http://arxiv.org/abs/2105.12757}
  {arXiv:2105.12757 [astro-ph.HE]} \BibitemShut {NoStop}%
\bibitem [{\citenamefont {Martinez}\ \emph {et~al.}(2022)\citenamefont
  {Martinez}, \citenamefont {Rodriguez},\ and\ \citenamefont
  {Fragione}}]{Martinez:2021tmr}%
  \BibitemOpen
  \bibfield  {author} {\bibinfo {author} {\bibfnamefont {M.~A.~S.}\
  \bibnamefont {Martinez}}, \bibinfo {author} {\bibfnamefont {C.~L.}\
  \bibnamefont {Rodriguez}}, \ and\ \bibinfo {author} {\bibfnamefont
  {G.}~\bibnamefont {Fragione}},\ }\href {\doibase 10.3847/1538-4357/ac8d55}
  {\bibfield  {journal} {\bibinfo  {journal} {Astrophys. J.}\ }\textbf
  {\bibinfo {volume} {937}},\ \bibinfo {pages} {78} (\bibinfo {year} {2022})},\
  \Eprint {http://arxiv.org/abs/2105.01671} {arXiv:2105.01671 [astro-ph.SR]}
  \BibitemShut {NoStop}%
\bibitem [{\citenamefont {Ryu}\ \emph {et~al.}(2022)\citenamefont {Ryu},
  \citenamefont {Perna},\ and\ \citenamefont {Wang}}]{Ryu:2022qpo}%
  \BibitemOpen
  \bibfield  {author} {\bibinfo {author} {\bibfnamefont {T.}~\bibnamefont
  {Ryu}}, \bibinfo {author} {\bibfnamefont {R.}~\bibnamefont {Perna}}, \ and\
  \bibinfo {author} {\bibfnamefont {Y.}~\bibnamefont {Wang}},\ }\href {\doibase
  10.1093/mnras/stac2316} {\bibfield  {journal} {\bibinfo  {journal} {Mon. Not.
  Roy. Astron. Soc.}\ }\textbf {\bibinfo {volume} {516}},\ \bibinfo {pages}
  {2204} (\bibinfo {year} {2022})},\ \Eprint {http://arxiv.org/abs/2206.00603}
  {arXiv:2206.00603 [astro-ph.HE]} \BibitemShut {NoStop}%
\bibitem [{\citenamefont {Campanelli}\ \emph {et~al.}(2008)\citenamefont
  {Campanelli}, \citenamefont {Lousto},\ and\ \citenamefont
  {Zlochower}}]{Campanelli:2007ea}%
  \BibitemOpen
  \bibfield  {author} {\bibinfo {author} {\bibfnamefont {M.}~\bibnamefont
  {Campanelli}}, \bibinfo {author} {\bibfnamefont {C.~O.}\ \bibnamefont
  {Lousto}}, \ and\ \bibinfo {author} {\bibfnamefont {Y.}~\bibnamefont
  {Zlochower}},\ }\href@noop {} {\bibfield  {journal} {\bibinfo  {journal}
  {Phys. Rev.}\ }\textbf {\bibinfo {volume} {D77}},\ \bibinfo {pages}
  {101501(R)} (\bibinfo {year} {2008})},\ \Eprint
  {http://arxiv.org/abs/0710.0879} {arXiv:0710.0879 [gr-qc]} \BibitemShut
  {NoStop}%
\bibitem [{\citenamefont {Lousto}\ and\ \citenamefont
  {Zlochower}(2008)}]{Lousto:2007rj}%
  \BibitemOpen
  \bibfield  {author} {\bibinfo {author} {\bibfnamefont {C.~O.}\ \bibnamefont
  {Lousto}}\ and\ \bibinfo {author} {\bibfnamefont {Y.}~\bibnamefont
  {Zlochower}},\ }\href {\doibase 10.1103/PhysRevD.77.024034} {\bibfield
  {journal} {\bibinfo  {journal} {Phys. Rev.}\ }\textbf {\bibinfo {volume}
  {D77}},\ \bibinfo {pages} {024034} (\bibinfo {year} {2008})},\ \Eprint
  {http://arxiv.org/abs/0711.1165} {arXiv:0711.1165 [gr-qc]} \BibitemShut
  {NoStop}%
\bibitem [{\citenamefont {Lousto}\ and\ \citenamefont
  {Nakano}(2008)}]{Lousto:2007ji}%
  \BibitemOpen
  \bibfield  {author} {\bibinfo {author} {\bibfnamefont {C.~O.}\ \bibnamefont
  {Lousto}}\ and\ \bibinfo {author} {\bibfnamefont {H.}~\bibnamefont
  {Nakano}},\ }\href@noop {} {\bibfield  {journal} {\bibinfo  {journal} {Class.
  Quant. Grav.}\ }\textbf {\bibinfo {volume} {25}},\ \bibinfo {pages} {195019}
  (\bibinfo {year} {2008})},\ \Eprint {http://arxiv.org/abs/arXiv:0710.5542
  [gr-qc]} {arXiv:0710.5542 [gr-qc]} \BibitemShut {NoStop}%
\bibitem [{\citenamefont {Laguna}(2004)}]{Laguna:2003sr}%
  \BibitemOpen
  \bibfield  {author} {\bibinfo {author} {\bibfnamefont {P.}~\bibnamefont
  {Laguna}},\ }\href@noop {} {\bibfield  {journal} {\bibinfo  {journal} {Phys.
  Rev. D}\ }\textbf {\bibinfo {volume} {69}},\ \bibinfo {pages} {104020}
  (\bibinfo {year} {2004})},\ \Eprint {http://arxiv.org/abs/gr-qc/0310073}
  {gr-qc/0310073} \BibitemShut {NoStop}%
\bibitem [{\citenamefont {Galaviz}\ \emph {et~al.}(2010)\citenamefont
  {Galaviz}, \citenamefont {Bruegmann},\ and\ \citenamefont
  {Cao}}]{Galaviz:2010mx}%
  \BibitemOpen
  \bibfield  {author} {\bibinfo {author} {\bibfnamefont {P.}~\bibnamefont
  {Galaviz}}, \bibinfo {author} {\bibfnamefont {B.}~\bibnamefont {Bruegmann}},
  \ and\ \bibinfo {author} {\bibfnamefont {Z.}~\bibnamefont {Cao}},\ }\href
  {\doibase 10.1103/PhysRevD.82.024005} {\bibfield  {journal} {\bibinfo
  {journal} {Phys. Rev. D}\ }\textbf {\bibinfo {volume} {82}},\ \bibinfo
  {pages} {024005} (\bibinfo {year} {2010})},\ \Eprint
  {http://arxiv.org/abs/1004.1353} {arXiv:1004.1353 [gr-qc]} \BibitemShut
  {NoStop}%
\bibitem [{\citenamefont {Bowen}\ and\ \citenamefont
  {York}(1980)}]{Bowen:1980yu}%
  \BibitemOpen
  \bibfield  {author} {\bibinfo {author} {\bibfnamefont {J.~M.}\ \bibnamefont
  {Bowen}}\ and\ \bibinfo {author} {\bibfnamefont {J.~W.}\ \bibnamefont {York},
  \bibfnamefont {Jr.}},\ }\href {\doibase 10.1103/PhysRevD.21.2047} {\bibfield
  {journal} {\bibinfo  {journal} {Phys. Rev.}\ }\textbf {\bibinfo {volume}
  {D21}},\ \bibinfo {pages} {2047} (\bibinfo {year} {1980})}\BibitemShut
  {NoStop}%
\bibitem [{\citenamefont {Zlochower}\ \emph {et~al.}(2005)\citenamefont
  {Zlochower}, \citenamefont {Baker}, \citenamefont {Campanelli},\ and\
  \citenamefont {Lousto}}]{Zlochower:2005bj}%
  \BibitemOpen
  \bibfield  {author} {\bibinfo {author} {\bibfnamefont {Y.}~\bibnamefont
  {Zlochower}}, \bibinfo {author} {\bibfnamefont {J.~G.}\ \bibnamefont
  {Baker}}, \bibinfo {author} {\bibfnamefont {M.}~\bibnamefont {Campanelli}}, \
  and\ \bibinfo {author} {\bibfnamefont {C.~O.}\ \bibnamefont {Lousto}},\
  }\href {\doibase 10.1103/PhysRevD.72.024021} {\bibfield  {journal} {\bibinfo
  {journal} {Phys. Rev.}\ }\textbf {\bibinfo {volume} {D72}},\ \bibinfo {pages}
  {024021} (\bibinfo {year} {2005})},\ \Eprint
  {http://arxiv.org/abs/gr-qc/0505055} {arXiv:gr-qc/0505055} \BibitemShut
  {NoStop}%
\bibitem [{\citenamefont {Brandt}\ and\ \citenamefont
  {Br{\"u}gmann}(1997)}]{Brandt97b}%
  \BibitemOpen
  \bibfield  {author} {\bibinfo {author} {\bibfnamefont {S.}~\bibnamefont
  {Brandt}}\ and\ \bibinfo {author} {\bibfnamefont {B.}~\bibnamefont
  {Br{\"u}gmann}},\ }\href@noop {} {\bibfield  {journal} {\bibinfo  {journal}
  {Phys. Rev. Lett.}\ }\textbf {\bibinfo {volume} {78}},\ \bibinfo {pages}
  {3606} (\bibinfo {year} {1997})},\ \Eprint
  {http://arxiv.org/abs/gr-qc/9703066} {gr-qc/9703066} \BibitemShut {NoStop}%
\bibitem [{\citenamefont {Ansorg}\ \emph {et~al.}(2004)\citenamefont {Ansorg},
  \citenamefont {Br\"ugmann},\ and\ \citenamefont {Tichy}}]{Ansorg:2004ds}%
  \BibitemOpen
  \bibfield  {author} {\bibinfo {author} {\bibfnamefont {M.}~\bibnamefont
  {Ansorg}}, \bibinfo {author} {\bibfnamefont {B.}~\bibnamefont {Br\"ugmann}},
  \ and\ \bibinfo {author} {\bibfnamefont {W.}~\bibnamefont {Tichy}},\
  }\href@noop {} {\bibfield  {journal} {\bibinfo  {journal} {Phys. Rev.}\
  }\textbf {\bibinfo {volume} {D70}},\ \bibinfo {pages} {064011} (\bibinfo
  {year} {2004})},\ \Eprint {http://arxiv.org/abs/gr-qc/0404056}
  {gr-qc/0404056} \BibitemShut {NoStop}%
\bibitem [{\citenamefont {Thornburg}(2004)}]{Thornburg2003:AH-finding}%
  \BibitemOpen
  \bibfield  {author} {\bibinfo {author} {\bibfnamefont {J.}~\bibnamefont
  {Thornburg}},\ }\href {\doibase 10.1088/0264-9381/21/2/026} {\bibfield
  {journal} {\bibinfo  {journal} {Class. Quant. Grav.}\ }\textbf {\bibinfo
  {volume} {21}},\ \bibinfo {pages} {743} (\bibinfo {year} {2004})},\ \Eprint
  {http://arxiv.org/abs/gr-qc/0306056} {gr-qc/0306056} \BibitemShut {NoStop}%
\bibitem [{\citenamefont {Campanelli}\ \emph {et~al.}(2007)\citenamefont
  {Campanelli}, \citenamefont {Lousto}, \citenamefont {Zlochower},
  \citenamefont {Krishnan},\ and\ \citenamefont {Merritt}}]{Campanelli:2006fy}%
  \BibitemOpen
  \bibfield  {author} {\bibinfo {author} {\bibfnamefont {M.}~\bibnamefont
  {Campanelli}}, \bibinfo {author} {\bibfnamefont {C.~O.}\ \bibnamefont
  {Lousto}}, \bibinfo {author} {\bibfnamefont {Y.}~\bibnamefont {Zlochower}},
  \bibinfo {author} {\bibfnamefont {B.}~\bibnamefont {Krishnan}}, \ and\
  \bibinfo {author} {\bibfnamefont {D.}~\bibnamefont {Merritt}},\ }\href@noop
  {} {\bibfield  {journal} {\bibinfo  {journal} {Phys. Rev.}\ }\textbf
  {\bibinfo {volume} {D75}},\ \bibinfo {pages} {064030} (\bibinfo {year}
  {2007})},\ \Eprint {http://arxiv.org/abs/gr-qc/0612076} {gr-qc/0612076}
  \BibitemShut {NoStop}%
\bibitem [{\citenamefont {Schnetter}\ \emph {et~al.}(2004)\citenamefont
  {Schnetter}, \citenamefont {Hawley},\ and\ \citenamefont
  {Hawke}}]{Schnetter-etal-03b}%
  \BibitemOpen
  \bibfield  {author} {\bibinfo {author} {\bibfnamefont {E.}~\bibnamefont
  {Schnetter}}, \bibinfo {author} {\bibfnamefont {S.~H.}\ \bibnamefont
  {Hawley}}, \ and\ \bibinfo {author} {\bibfnamefont {I.}~\bibnamefont
  {Hawke}},\ }\href@noop {} {\bibfield  {journal} {\bibinfo  {journal} {Class.
  Quant. Grav.}\ }\textbf {\bibinfo {volume} {21}},\ \bibinfo {pages} {1465}
  (\bibinfo {year} {2004})},\ \Eprint {http://arxiv.org/abs/gr-qc/0310042}
  {gr-qc/0310042} \BibitemShut {NoStop}%
\bibitem [{\citenamefont {Campanelli}\ \emph {et~al.}(2009)\citenamefont
  {Campanelli}, \citenamefont {Lousto}, \citenamefont {Nakano},\ and\
  \citenamefont {Zlochower}}]{Campanelli:2008nk}%
  \BibitemOpen
  \bibfield  {author} {\bibinfo {author} {\bibfnamefont {M.}~\bibnamefont
  {Campanelli}}, \bibinfo {author} {\bibfnamefont {C.~O.}\ \bibnamefont
  {Lousto}}, \bibinfo {author} {\bibfnamefont {H.}~\bibnamefont {Nakano}}, \
  and\ \bibinfo {author} {\bibfnamefont {Y.}~\bibnamefont {Zlochower}},\ }\href
  {\doibase 10.1103/PhysRevD.79.084010} {\bibfield  {journal} {\bibinfo
  {journal} {Phys. Rev.}\ }\textbf {\bibinfo {volume} {D79}},\ \bibinfo {pages}
  {084010} (\bibinfo {year} {2009})},\ \Eprint {http://arxiv.org/abs/0808.0713}
  {arXiv:0808.0713 [gr-qc]} \BibitemShut {NoStop}%
\bibitem [{\citenamefont {Alcubierre}\ \emph {et~al.}(2005)\citenamefont
  {Alcubierre} \emph {et~al.}}]{Alcubierre:2004hr}%
  \BibitemOpen
  \bibfield  {author} {\bibinfo {author} {\bibfnamefont {M.}~\bibnamefont
  {Alcubierre}} \emph {et~al.},\ }\href@noop {} {\bibfield  {journal} {\bibinfo
   {journal} {Phys. Rev.}\ }\textbf {\bibinfo {volume} {D72}},\ \bibinfo
  {pages} {044004} (\bibinfo {year} {2005})},\ \Eprint
  {http://arxiv.org/abs/gr-qc/0411149} {gr-qc/0411149} \BibitemShut {NoStop}%
\bibitem [{\citenamefont {Campanelli}\ and\ \citenamefont
  {Lousto}(1999)}]{Campanelli:1998jv}%
  \BibitemOpen
  \bibfield  {author} {\bibinfo {author} {\bibfnamefont {M.}~\bibnamefont
  {Campanelli}}\ and\ \bibinfo {author} {\bibfnamefont {C.~O.}\ \bibnamefont
  {Lousto}},\ }\href {\doibase 10.1103/PhysRevD.59.124022} {\bibfield
  {journal} {\bibinfo  {journal} {Phys. Rev.}\ }\textbf {\bibinfo {volume}
  {D59}},\ \bibinfo {pages} {124022} (\bibinfo {year} {1999})},\ \Eprint
  {http://arxiv.org/abs/gr-qc/9811019} {arXiv:gr-qc/9811019 [gr-qc]}
  \BibitemShut {NoStop}%
\bibitem [{\citenamefont {Lousto}\ and\ \citenamefont
  {Zlochower}(2007)}]{Lousto:2007mh}%
  \BibitemOpen
  \bibfield  {author} {\bibinfo {author} {\bibfnamefont {C.~O.}\ \bibnamefont
  {Lousto}}\ and\ \bibinfo {author} {\bibfnamefont {Y.}~\bibnamefont
  {Zlochower}},\ }\href@noop {} {\bibfield  {journal} {\bibinfo  {journal}
  {Phys. Rev.}\ }\textbf {\bibinfo {volume} {D76}},\ \bibinfo {pages}
  {041502(R)} (\bibinfo {year} {2007})},\ \Eprint
  {http://arxiv.org/abs/gr-qc/0703061} {gr-qc/0703061} \BibitemShut {NoStop}%
\bibitem [{\citenamefont {Healy}\ \emph {et~al.}(2017)\citenamefont {Healy},
  \citenamefont {Lousto}, \citenamefont {Nakano},\ and\ \citenamefont
  {Zlochower}}]{Healy:2017zqj}%
  \BibitemOpen
  \bibfield  {author} {\bibinfo {author} {\bibfnamefont {J.}~\bibnamefont
  {Healy}}, \bibinfo {author} {\bibfnamefont {C.~O.}\ \bibnamefont {Lousto}},
  \bibinfo {author} {\bibfnamefont {H.}~\bibnamefont {Nakano}}, \ and\ \bibinfo
  {author} {\bibfnamefont {Y.}~\bibnamefont {Zlochower}},\ }\href {\doibase
  10.1088/1361-6382/aa7929} {\bibfield  {journal} {\bibinfo  {journal} {Class.
  Quant. Grav.}\ }\textbf {\bibinfo {volume} {34}},\ \bibinfo {pages} {145011}
  (\bibinfo {year} {2017})},\ \Eprint {http://arxiv.org/abs/1702.00872}
  {arXiv:1702.00872 [gr-qc]} \BibitemShut {NoStop}%
\bibitem [{\citenamefont {Woodford}\ \emph {et~al.}(2019)\citenamefont
  {Woodford}, \citenamefont {Boyle},\ and\ \citenamefont
  {Pfeiffer}}]{Woodford:2019tlo}%
  \BibitemOpen
  \bibfield  {author} {\bibinfo {author} {\bibfnamefont {C.~J.}\ \bibnamefont
  {Woodford}}, \bibinfo {author} {\bibfnamefont {M.}~\bibnamefont {Boyle}}, \
  and\ \bibinfo {author} {\bibfnamefont {H.~P.}\ \bibnamefont {Pfeiffer}},\
  }\href {\doibase 10.1103/PhysRevD.100.124010} {\bibfield  {journal} {\bibinfo
   {journal} {Phys. Rev.}\ }\textbf {\bibinfo {volume} {D100}},\ \bibinfo
  {pages} {124010} (\bibinfo {year} {2019})},\ \Eprint
  {http://arxiv.org/abs/1904.04842} {arXiv:1904.04842 [gr-qc]} \BibitemShut
  {NoStop}%
\bibitem [{\citenamefont {Healy}\ and\ \citenamefont
  {Lousto}(2020)}]{Healy:2020vre}%
  \BibitemOpen
  \bibfield  {author} {\bibinfo {author} {\bibfnamefont {J.}~\bibnamefont
  {Healy}}\ and\ \bibinfo {author} {\bibfnamefont {C.~O.}\ \bibnamefont
  {Lousto}},\ }\href {\doibase 10.1103/PhysRevD.102.104018} {\bibfield
  {journal} {\bibinfo  {journal} {Phys. Rev. D}\ }\textbf {\bibinfo {volume}
  {102}},\ \bibinfo {pages} {104018} (\bibinfo {year} {2020})},\ \Eprint
  {http://arxiv.org/abs/2007.07910} {arXiv:2007.07910 [gr-qc]} \BibitemShut
  {NoStop}%
\bibitem [{\citenamefont {Lousto}\ \emph {et~al.}(2016)\citenamefont {Lousto},
  \citenamefont {Healy},\ and\ \citenamefont {Nakano}}]{Lousto:2015uwa}%
  \BibitemOpen
  \bibfield  {author} {\bibinfo {author} {\bibfnamefont {C.~O.}\ \bibnamefont
  {Lousto}}, \bibinfo {author} {\bibfnamefont {J.}~\bibnamefont {Healy}}, \
  and\ \bibinfo {author} {\bibfnamefont {H.}~\bibnamefont {Nakano}},\ }\href
  {\doibase 10.1103/PhysRevD.93.044031} {\bibfield  {journal} {\bibinfo
  {journal} {Phys. Rev.}\ }\textbf {\bibinfo {volume} {D93}},\ \bibinfo {pages}
  {044031} (\bibinfo {year} {2016})},\ \Eprint
  {http://arxiv.org/abs/1506.04768} {arXiv:1506.04768 [gr-qc]} \BibitemShut
  {NoStop}%
\bibitem [{\citenamefont {Peters}(1964)}]{Peters:1964}%
  \BibitemOpen
  \bibfield  {author} {\bibinfo {author} {\bibfnamefont {P.~C.}\ \bibnamefont
  {Peters}},\ }\href@noop {} {\bibfield  {journal} {\bibinfo  {journal} {Phys.
  Rev.}\ }\textbf {\bibinfo {volume} {136}},\ \bibinfo {pages} {B1224}
  (\bibinfo {year} {1964})}\BibitemShut {NoStop}%
\bibitem [{\citenamefont {Datta}(2023)}]{Datta:2023uln}%
  \BibitemOpen
  \bibfield  {author} {\bibinfo {author} {\bibfnamefont {S.}~\bibnamefont
  {Datta}},\ }\href@noop {} {\  (\bibinfo {year} {2023})},\ \Eprint
  {http://arxiv.org/abs/2306.12522} {arXiv:2306.12522 [gr-qc]} \BibitemShut
  {NoStop}%
\bibitem [{\citenamefont {Lousto}\ and\ \citenamefont
  {Healy}(2019)}]{Lousto:2019lyf}%
  \BibitemOpen
  \bibfield  {author} {\bibinfo {author} {\bibfnamefont {C.~O.}\ \bibnamefont
  {Lousto}}\ and\ \bibinfo {author} {\bibfnamefont {J.}~\bibnamefont {Healy}},\
  }\href {\doibase 10.1103/PhysRevD.100.104039} {\bibfield  {journal} {\bibinfo
   {journal} {Phys. Rev.}\ }\textbf {\bibinfo {volume} {D100}},\ \bibinfo
  {pages} {104039} (\bibinfo {year} {2019})},\ \Eprint
  {http://arxiv.org/abs/1908.04382} {arXiv:1908.04382 [gr-qc]} \BibitemShut
  {NoStop}%
\bibitem [{\citenamefont {Bai}\ \emph {et~al.}(2011)\citenamefont {Bai},
  \citenamefont {Cao}, \citenamefont {Han}, \citenamefont {Lin}, \citenamefont
  {Yo},\ and\ \citenamefont {Yu}}]{Bai:2011za}%
  \BibitemOpen
  \bibfield  {author} {\bibinfo {author} {\bibfnamefont {S.}~\bibnamefont
  {Bai}}, \bibinfo {author} {\bibfnamefont {Z.-J.}\ \bibnamefont {Cao}},
  \bibinfo {author} {\bibfnamefont {W.-B.}\ \bibnamefont {Han}}, \bibinfo
  {author} {\bibfnamefont {C.-Y.}\ \bibnamefont {Lin}}, \bibinfo {author}
  {\bibfnamefont {H.-J.}\ \bibnamefont {Yo}}, \ and\ \bibinfo {author}
  {\bibfnamefont {J.-P.}\ \bibnamefont {Yu}},\ }\href {\doibase
  10.1088/1742-6596/330/1/012016} {\bibfield  {journal} {\bibinfo  {journal}
  {J. Phys. Conf. Ser.}\ }\textbf {\bibinfo {volume} {330}},\ \bibinfo {pages}
  {012016} (\bibinfo {year} {2011})},\ \Eprint {http://arxiv.org/abs/1203.6185}
  {arXiv:1203.6185 [gr-qc]} \BibitemShut {NoStop}%
\bibitem [{\citenamefont {Imbrogno}\ \emph {et~al.}(2023)\citenamefont
  {Imbrogno}, \citenamefont {Meringolo},\ and\ \citenamefont
  {Servidio}}]{Imbrogno:2021xrh}%
  \BibitemOpen
  \bibfield  {author} {\bibinfo {author} {\bibfnamefont {M.}~\bibnamefont
  {Imbrogno}}, \bibinfo {author} {\bibfnamefont {C.}~\bibnamefont {Meringolo}},
  \ and\ \bibinfo {author} {\bibfnamefont {S.}~\bibnamefont {Servidio}},\
  }\href {\doibase 10.1088/1361-6382/acb881} {\bibfield  {journal} {\bibinfo
  {journal} {Class. Quant. Grav.}\ }\textbf {\bibinfo {volume} {40}},\ \bibinfo
  {pages} {075008} (\bibinfo {year} {2023})},\ \Eprint
  {http://arxiv.org/abs/2108.01392} {arXiv:2108.01392 [gr-qc]} \BibitemShut
  {NoStop}%
\bibitem [{\citenamefont {Flanagan}\ and\ \citenamefont
  {Hinderer}(2008)}]{Flanagan:2007ix}%
  \BibitemOpen
  \bibfield  {author} {\bibinfo {author} {\bibfnamefont {E.~E.}\ \bibnamefont
  {Flanagan}}\ and\ \bibinfo {author} {\bibfnamefont {T.}~\bibnamefont
  {Hinderer}},\ }\href {\doibase 10.1103/PhysRevD.77.021502} {\bibfield
  {journal} {\bibinfo  {journal} {Phys. Rev.}\ }\textbf {\bibinfo {volume}
  {D77}},\ \bibinfo {pages} {021502} (\bibinfo {year} {2008})},\ \Eprint
  {http://arxiv.org/abs/0709.1915} {arXiv:0709.1915 [astro-ph]} \BibitemShut
  {NoStop}%
\bibitem [{\citenamefont {Galaviz}\ and\ \citenamefont
  {Bruegmann}(2011)}]{Galaviz:2010te}%
  \BibitemOpen
  \bibfield  {author} {\bibinfo {author} {\bibfnamefont {P.}~\bibnamefont
  {Galaviz}}\ and\ \bibinfo {author} {\bibfnamefont {B.}~\bibnamefont
  {Bruegmann}},\ }\href {\doibase 10.1103/PhysRevD.83.084013} {\bibfield
  {journal} {\bibinfo  {journal} {Phys. Rev.}\ }\textbf {\bibinfo {volume}
  {D83}},\ \bibinfo {pages} {084013} (\bibinfo {year} {2011})},\ \Eprint
  {http://arxiv.org/abs/1012.4423} {arXiv:1012.4423 [gr-qc]} \BibitemShut
  {NoStop}%
\bibitem [{\citenamefont {Naoz}\ \emph {et~al.}(2013)\citenamefont {Naoz},
  \citenamefont {Kocsis}, \citenamefont {Loeb},\ and\ \citenamefont
  {Yunes}}]{Naoz:2012bx}%
  \BibitemOpen
  \bibfield  {author} {\bibinfo {author} {\bibfnamefont {S.}~\bibnamefont
  {Naoz}}, \bibinfo {author} {\bibfnamefont {B.}~\bibnamefont {Kocsis}},
  \bibinfo {author} {\bibfnamefont {A.}~\bibnamefont {Loeb}}, \ and\ \bibinfo
  {author} {\bibfnamefont {N.}~\bibnamefont {Yunes}},\ }\href {\doibase
  10.1088/0004-637X/773/2/187} {\bibfield  {journal} {\bibinfo  {journal}
  {Astrophys. J.}\ }\textbf {\bibinfo {volume} {773}},\ \bibinfo {pages} {187}
  (\bibinfo {year} {2013})},\ \Eprint {http://arxiv.org/abs/1206.4316}
  {arXiv:1206.4316 [astro-ph.SR]} \BibitemShut {NoStop}%
\end{thebibliography}%

\end{document}